\documentclass{amsart}
\usepackage[dvips]{epsfig}
\usepackage{graphicx}
\usepackage{latexsym}
\usepackage{amsmath}
\usepackage{amsthm}
\usepackage{amssymb}
\usepackage{mathrsfs}
\usepackage[shortlabels]{enumitem}
\newlist{propenum}{enumerate}{1} 
\setlist[propenum]{label=(\roman*)}

\RequirePackage[colorlinks=true,citecolor=blue,urlcolor=blue]{hyperref}

\usepackage{caption}
\usepackage{subcaption}

\usepackage{eucal}

\usepackage{xcolor,soul}

\usepackage{mathtools}

 \usepackage[applemac]{inputenc}
\usepackage{fge}
\newcommand{\mysetminus}{\mathbin{\fgebackslash}}

\newtheorem{thm}{Theorem}[section]

\newtheorem{assumption}[thm]{Assumption}
\newtheorem{lem}[thm]{Lemma}
\newtheorem{cor}[thm]{Corollary}
\newtheorem{defi}[thm]{Definition}
\newtheorem{hyp}[thm]{Assumption}

\theoremstyle{remark}

\newtheorem{rem}[thm]{Remark}

\newcommand{\vip}{\vskip.2cm}
\newcommand{\COMMENTAIRE}[1]{}

\newcommand{\field}[1]{\mathbb{#1}}

\newcommand{\GG}{\field{G}}

\newcommand{\NN}{\field{N}}
\newcommand{\PP}{\field{P}}

\newcommand{\RR}{\field{R}}
\newcommand{\TT}{\field{T}}

\newcommand{\Bb}{{\mathcal B}}

\newcommand{\Ff}{{\mathcal F}}
\newcommand{\Gg}{{\mathcal G}}
\newcommand{\Hh}{{\mathcal H}}
\newcommand{\Ll}{{\mathcal L}}

\newcommand{\Nn}{{\mathcal N}}
\newcommand{\Pp}{{\mathcal P}}
\newcommand{\Rr}{{\mathcal R}}

\newcommand{\Tt}{{\mathcal T}}

\newcommand{\Qq}{{\mathcal Q}}
\newcommand{\vt}{{\vartriangle}}

\def \ep {\varepsilon}

\newcommand{\A}{{\mathbb A}}

\newcommand{\E}{{\mathbb E}}

\newcommand{\ind}{{\bf 1}}

\DeclareMathOperator*{\argmin}{arg\,min} 
\DeclareMathOperator\sinc{sinc}

\newcommand{\bigO}{\mathcal{O}}

\newcommand\smallO{
  \mathchoice
    {{\scriptstyle\mathcal{O}}}
    {{\scriptstyle\mathcal{O}}}
    {{\scriptscriptstyle\mathcal{O}}}
    {\scalebox{.7}{$\scriptscriptstyle\mathcal{O}$}}
  }

\begin{document}

\title[LS method for BMC]{Least squares estimation of the transition density in bifurcating Markov 
models
}

\author{S. Val\`ere Bitseki Penda}

\address{S. Val\`ere Bitseki Penda, Universit\'e Bourgogne Europe, CNRS, IMB UMR 5584, F-21000 Dijon, France.}

\email{simeon-valere.bitseki-penda@u-bourgogne.fr}

\date{\today}

\begin{abstract}
In this article, we propose a least squares method for the estimation of the transition density in bifurcating Markov models. Unlike the kernel estimation, this method do not use the quotient which can be a source of errors. In order to study the rate of convergence for least squares estimators, we develop exponential inequalities for empirical process of bifurcating Markov chain under bracketing assumption. Unlike the classical processes, we observe that for bifurcating Markov chains, the complexity parameter depends on the ergodicity rate and as consequence, we have that the convergence rate of our estimator is a function of the ergodicity rate. We conclude with a numerical study to validate our theoretical results.     
\end{abstract}

\maketitle





\section{Introduction}
First, we give a precise definition of bifurcating Markov chain. In the sequel, $\RR$ will denote the set of real numbers and for $p \geq 1$, $\RR^{p}$ will denote the set of vectors with $p$ real components. For all $p \geq 1$, we will equip $\RR^{p}$ with the usual $\sigma\text{-}$field $\Bb(\RR^{p})$.

\subsection{The binary tree associated to bifurcating Markov chain}

We denote by $\NN$ the set of natural integers. For $m \in \NN$, we set $\GG_{m} = \{0,1\}^{m}$, the $m\text{-}$times cartesian product of the set $\{0,1\}$, with $\GG_{0} = \{\emptyset\}.$ We consider the binary tree $\TT = \bigcup_{m = 0}^{+\infty} \GG_{m}.$ For $u \in \GG_{m},$ we set $|u| = m$ and define the concatenation $u0 = (u,0) \in \GG_{m+1}$ and $u1 = (u,1) \in \GG_{m+1}.$ One can seen $\TT$ as a population in which each individual $u \in \TT$ as two offspring: one of type $0$, $u0$ and the other of type 1,  $u1.$ Therefore, for $m \in \NN$, $\GG_{m}$ denotes the $m\text{-th}$ generation of the population and for $u \in \TT$, $|u|$ is the generation of the individual $u$. 

\subsection{Bifurcating Markov chain}

Let $\Pp$ be a transition probability from $(\RR,\Bb(\RR))$ to $(\RR^{2},\Bb(\RR^{2}))$, that is:
\begin{itemize}
\item $\Pp(\cdot,A)$ is measurable function for all $A \in \Bb(\RR^{2});$
\item $\Pp(x,\cdot)$ is a probability measure on $(\RR^{2}, \Bb(\RR^{2}))$ for all $x \in \RR.$
\end{itemize}
Let $\nu$ be a probability measure on $(\RR, \Bb(\RR))$ and $(\Omega,\Ff,(\Ff_{m})_{m\geq 0},\PP)$ be a filtered probability space. 
\begin{defi}\label{def:bmc}
We say that a family of $\RR\text{-}$valued random variables $(X_{u})_{u \in \TT}$ defined on \, \, \, $(\Omega,\Ff,(\Ff_{m})_{m\geq 0},\PP)$ is a bifurcating Markov chain with initial probability $\nu$ and transition probability $\Pp$ if:
\begin{itemize}
\item $X_{\emptyset}$ is distributed as $\nu$;
\item $X_{u}$ is $\Ff_{|u|}\text{-}$measurable for all $u \in \TT$;
\item for every $m \geq 0$ and any family of (bounded) measurable functions $(g_u)_{u \in \mathbb G_m},$
\begin{equation*}
\E\big[\prod_{u \in \mathbb G_m} g_u(X_u, X_{u0}, X_{u1})\big|\mathcal F_m\big]= \prod_{u \in \mathbb G_m}\Pp g_u(X_u),
\end{equation*}
where
$\Pp g(x)=\int_{\RR^{2}}g(x,y,z) \Pp(x,dy\,dz)$ denotes the action of $\Pp$ on $g$.
\end{itemize}
\end{defi}

\subsection{Useful quantities associated to the study of bifurcating Markov chains}

Let $(X_{u})_{u \in \TT}$ be a $\RR\text{-}$valued bifurcating Markov chain with initial probability $\nu$ and transition probability $\Pp.$ For a finite subset $\A$ of $\TT$, we denote by $|\A|$ the cardinality of $\A$. We consider the following sums: 
\begin{equation*}
\begin{cases}
M_{\A}(f) = \sum_{u \in \A} f(X_{u}) & \text{if $f$ is defined on $\RR$} \\
M_{\A}(f) = \sum_{u \in \A} f(X_{u},X_{u0},X_{u1}) & \text{if $f$ is defined on $\RR^{3}$}.
\end{cases}
\end{equation*}
In this paper, we are particularly interested to the case $\A = \GG_{n}$, for $n \in \NN.$ We have in that case, $|\GG_{n}| = 2^{n}.$
We denote by $\Pp_{0}$ and $\Pp_{1}$ the first and the second marginal of $\Pp$, that is, for all $x \in \RR$ and $A \subset \RR,$ $\Pp_{0}(x,A) = \Pp(x,A\times \RR)$ and $\Pp_{1}(x,A) = \Pp(x,\RR \times A).$ We set $\Qq = \tfrac{1}{2}(\Pp_{0} + \Pp_{1}).$ Then $\Qq$ is a transition probability on $(\RR, \Bb(\RR)).$ We denote by $(Y_{m})_{m \in \NN}$ the Markov chain with transition probability $\Qq$. The Markov chain $(Y_{m})_{m \in \NN}$, that we call the tagged chain, corresponds to a lineage taken at random in $(X_{u})_{u \in \TT}$. In the theory of bifurcating Markov chains, $(Y_{m})_{m\in \NN}$ plays a key role for both asymptotic and non-asymptotic analysis of bifurcating Markov chains (see \cite{BDG14, BHO, Guyon}). In particular, Guyon \cite{Guyon} prove that if $(Y_{m})_{m \in \NN}$ is ergodic with invariant probability measure $\mu$, then for a test function $f$, the following convergence holds almost-surely:
\begin{equation*}
\begin{cases}
\lim_{n \rightarrow \infty} |\GG_{n}|^{-1}M_{\GG_{n}}(f) = \langle \mu, f\rangle & \text{if $f$ is defined on $\RR$}\\
\lim_{n \rightarrow \infty} |\GG_{n}|^{-1}M_{\GG_{n}}(f) = \langle \mu, \Pp f\rangle & \text{if $f$ is defined on $\RR^{3},$}\
\end{cases}
\end{equation*}
where for a function $g$ defined on $\RR$, we set
\begin{equation*}
\langle \mu,g \rangle = \int_{\RR} g(x)\mu(dx).
\end{equation*}
In the sequel, we will assume that the Markov chain $(Y_{m})_{m \in \NN}$ is ergodic and we denote by $\mu$ its invariant probability measure.

\begin{rem}\label{rem:modau-trig}{({\bf The mother-daughter triangle process, see \cite{bitsekiS2024}})}
For all $u\in \TT$, we denote by $X_{u}^{\vartriangle} = (X_{u},X_{u0},X_{u1})$ the mother-daughter triangle. One can check that $(X_{u}^{\vartriangle},u\in\TT)$ is a bifurcating Markov chain on $\RR^{3}$ with transition probability $\Pp^{\vartriangle}$ defined by
\[
\Pp^{\vartriangle}(xx_{0}x_{1}, dyy_{0}y_{1}, dzz_{0}z_{1}) = \delta_{x_{0}}(dy)\Pp(y,dy_{0},dy_{1})\delta_{x_{1}}(dz)\Pp(z,dz_{0},dz_{1}),
\]
where we set  $xx_{0}x_{1} = (x,x_{0},x_{1}).$ 

The transition probability, $\Qq^{\vartriangle}$, of the tagged chain associated to $X^{\vartriangle}$  is defined by
\[
\Qq^{\vartriangle}(xx_{0}x_{1}, dyy_{0}y_{1}) = \frac{1}{2}(\delta_{x_{0}}(dy) + \delta_{x_{1}}(dy))\Pp(y,dy_{0},dy_{1}).
\]
If $\mu$ is the invariant probability measure of $\Qq$, then the probability measure $\mu^{\vartriangle}$ defined on $\RR^{3}$ by
\begin{equation*}
\mu^{\vartriangle}(dxx_{0}x_{1}) = \mu(dx)\Pp(x,dx_{0},dx_{1})
\end{equation*}
is the invariant probability measure  of $\Qq^{\vartriangle}$, that is, for all $f \in \Bb(\RR^{3})$, we have $\mu^{\vartriangle}\Qq^{\vartriangle}f = \langle \mu^{\vartriangle}, f \rangle$. One can also check the following: for all $n\in\NN^{*}$, we have
\begin{equation}\label{eq:Qtrin}
(\Qq^{\vartriangle})^{n}f = \frac{1}{2}(\Qq^{n-1}\Pp f \oplus \Qq^{n-1}\Pp f).
\end{equation}

\end{rem}

\subsection{A brief history of the study of bifurcating Markov chains models}
The bifurcating Markov chains models were introduced in the literature with the aim to understand the mechanisms of cell division. The first model of bifurcating Markov chain, named symmetric bifurcating auto-regressive process (BAR) were introduced by Cowan and Staudte \cite{CS86} in order to analyse the Escherichia Coli (E. Coli, in short) lineage data. E. Coli is a rod shaped bacterium which reproduce by dividing in two producing two cells: one of type 0, called the new pole, which has the new end of the mother and the other of type 1, called the old pole, which has the old end of the mother. After this work of Cowan and Staudte, the BAR model was intensively studied and many extensions were proposed. One can cite for exemple the work of  Basawa and Zhou \cite{BZ04} where the term bifurcating Markov chain appeared for the first time. But it was Guyon in \cite{Guyon} who developed a theory for BMC.  Guyon's work has in particular permit to prove statistical evidence of aging in Escherichia Coli. Recently, several works have been devoted to the study of nonparametric statistics associated with BMC models \cite{bitseki23, BHO, bitsekiroche2020, BO17}.

\medskip
From the definition \ref{def:bmc}, one can see that the distribution of $(X_{u})_{u \in \TT}$ is entirely determined by the initial law $\nu$ and the transition probability $\Pp.$ Actually, this transition probability is unknown and the aim of this paper  is to estimate it from the data. For that purpose, we do the following assumption.
\begin{hyp}\label{hyp:DenMu}
The transition kernel $\Pp$ has a density, still denoted by $\Pp$, with respect to the Lebesgue measure. 
\end{hyp}
\begin{rem}\label{rem:}
As a consequence of Assumption \ref{hyp:DenMu}, we have that the transition kernel $\Qq$ has a density, still denoted by $\Qq$, with respect to  the Lebesgue measure. More precisely, we have $\Qq(x,y) = 2^{-1} \int_{S} (\Pp(x,y,z)+\Pp(x,z,y))dz.$ This implies in particular that the invariant probability $\mu$ has a density, still denoted by $\mu$, with respect to the Lebesgue measure (for more details, we refer for e.g. to \cite{duflo2013random}, chap 6).
\end{rem}
In completion with Assumption \ref{hyp:DenMu}, we do the following assumption.
\begin{assumption} \label{ass:densityq}
The transition densities $\Pp$ and $\Qq$ and the invariant density $\mu$ are uniformly bounded, that is
\begin{equation*}
C_{0} = \sup_{x,y,z \in \RR}\{\mu(x) + \Qq(x,y) + \Pp(x,y,z)\} < \infty.
\end{equation*}
\end{assumption}

\section{Least squares estimators of the transition density $\Pp$}

We consider the basis of functions $\pmb{\varphi} = (\varphi_{1}, \ldots, \varphi_{d})^{t},$ where $\mathbf{v}^{t}$ denotes the transpose of the vector $\mathbf{v}$ and for all $j \in \{1, \ldots, d\}$, $\varphi_{j}$ is a nonnegative function defined on $\RR^{3}.$ We estimate the density $\Pp$ by
\begin{equation}\label{eq:wide-P}
\widehat{\Pp} = \sum_{j = 1}^{d} \beta^{*}_{j} \varphi_{j} = \pmb{\beta}^{*t} \pmb{\varphi}
\end{equation} 
where the scalar vector $\pmb{\beta}^{*} = (\beta^{*}_{1}, \ldots, \beta^{*}_{d})^{t}$ is unknown. In order to determine $\pmb{\beta}^{*},$ one minimize the functional $J_{0}$ defined, for all $\pmb{\beta} = (\beta_{1}, \ldots, \beta_{d})^{t}$, by
\begin{equation*}
J_{0}(\pmb{\beta}) = \frac{1}{2}\int_{\RR^{3}} ((\pmb{\beta}^{t} \pmb{\varphi})(x,y,z) - \Pp(x,y,z))^{2} \mu(x)dxdydz.
\end{equation*}
Using \eqref{eq:wide-P}, easy calculations lead us to
\begin{multline}\label{eq:J0}
J_{0}(\pmb{\beta}) = \frac{1}{2} \int_{\RR^{3}} (\pmb{\beta}^{t} \pmb{\varphi})^{2}(x,y,z) \mu(x)dxdydz - \int_{\RR^{3}} (\pmb{\beta}^{t} \pmb{\varphi})(x,y,z)\Pp(x,y,z)\mu(x)dxdydz  \\ 
+ \frac{1}{2}\int_{\RR^{3}} \Pp^{2}(x,y,z) \mu(x)dxdydz.
\end{multline}
Since $\pmb{\beta}$ do not appear in the last term of the right hand in \eqref{eq:J0}, then, to minimize $J_{0}$, it suffices to minimize the function $J$ defined by
\begin{equation*}
J(\pmb{\beta}) = \frac{1}{2} \int_{\RR^{3}} (\pmb{\beta}^{t} \pmb{\varphi})^{2}(x,y,z) \mu(x)dxdydz - \int_{\RR^{3}} (\pmb{\beta}^{t} \pmb{\varphi})(x,y,z)\Pp(x,y,z)\mu(x)dxdydz. 
\end{equation*}
We consider the matrix $\pmb{H} = (\pmb{H}_{i,j})_{1\leq i,j \leq d}$ and the vector $\pmb{h} = (\pmb{h}_{i})_{1\leq i \leq d}$ defined by
\begin{equation*}
\pmb{H}_{i,j} = \langle \mu, \overline{\varphi}_{i,j} \rangle \quad \text{and} \quad \pmb{h}_{i} = \langle \mu, \Pp \varphi_{i} \rangle,
\end{equation*}
where
\begin{equation}\label{eq:basis}
\overline{\varphi}_{i,j}(x) = \int_{\RR^{2}} (\varphi_{i} \varphi_{j})(x,y,z) dy dz; \, \langle \mu, \overline{\varphi}_{i,j} \rangle = \int_{\RR} \overline{\varphi}_{i,j}(x)\mu(x)dx; \, \langle \mu, \Pp \varphi_{i} \rangle = \int_{\RR} (\Pp \varphi_{i})(x) \mu(x) dx.
\end{equation}
Then we have
\begin{equation}\label{eq:def-J}
J(\pmb{\beta}) = \frac{1}{2}\pmb{\beta}^{t} \pmb{H} \pmb{\beta} - \pmb{\beta}^{t} \pmb{h}.
\end{equation}
The unknown parameter $\pmb{\beta}^{*}$ is given by
\begin{equation}\label{eq:optim1}
\pmb{\beta}^{*} = \argmin_{\pmb{\beta} \in \RR^{d}} J(\pmb{\beta}).
\end{equation}
Now, given that $\pmb{H}$ and $\pmb{h}$ in \eqref{eq:def-J} are unknown, since the invariant density $\mu$ and the transition density $\Pp$ are unknown, we will replace them by their estimates. As estimators for $\pmb{H}$ and $\pmb{h}$ respectively, we propose $\widehat{\pmb{H}} = (\widehat{\pmb{H}}_{i,j})_{1\leq i,j \leq d}$ and $\widehat{\pmb{h}} = (\widehat{\pmb{h}}_{i})_{1 \leq i \leq d}$ defined by:
\begin{equation*}
\widehat{\pmb{H}}_{i,j} = \frac{1}{|\GG_{n}|} \sum_{u \in \GG_{n}} \overline{\varphi}_{i,j}(X_{u}) \quad \text{and} \quad \widehat{\pmb{h}}_{i} = \frac{1}{|\GG_{n}|} \sum_{u \in \GG_{n}} \varphi_{j}(X_{u},X_{u0},X_{u1}).
\end{equation*} 
The theoretical explanation for using these estimators come from the work of Guyon \cite{Guyon}. Indeed, from Theorems 11 and 12 in \cite{Guyon}, we have under mild conditions that $\widehat{\pmb{H}}$ and $\widehat{\pmb{h}}$ are strongly consistent estimators of $\pmb{H}$ and $\pmb{h}$ respectively. The convergence rates of these kinds of estimators have been studied by Bitseki \& \textit{al.} \cite{BDG14} and the fluctuations have been studied recently by Bitseki and Delmas \cite{BD2,BD2020}.  We can now replace $J$ defined in \eqref{eq:def-J} by its estimate $\widehat{J}$ defined for all $\pmb{\beta} \in \RR^{d}$  by:
\begin{equation*}
\widehat{J}(\pmb{\beta}) = \frac{1}{2}\pmb{\beta}^{t}\widehat{\pmb{H}}\pmb{\beta} - \pmb{\beta}^{t} \widehat{\pmb{h}}.
\end{equation*}
Furthermore, for $\lambda > 0$, we replace the optimization problem \eqref{eq:optim1} by:
\begin{equation}\label{eq:optim2}
\widetilde{\pmb{\beta}} = \argmin_{\pmb{\beta} \in \RR^{d}} \{\widehat{J}(\pmb{\beta}) + \frac{\lambda}{2} \pmb{\beta}^{t} \pmb{\beta}\},
\end{equation} 
where for $\lambda > 0$, the regularizer $\frac{\lambda}{2} \pmb{\beta}^{t} \pmb{\beta}$ is introduced in order to stabilize the problem. Note that $\widetilde{\pmb{\beta}}$ is an estimate of $\pmb{\beta}^{*}$. We denote by $\pmb{I}_{d}$ the identity matrix of size $d$. Then, after simple calculations, we have
\begin{equation*}
\widetilde{\pmb{\beta}} = (\widehat{\pmb{H}} + \lambda \pmb{I}_{d})^{-1} \widehat{\pmb{h}}.
\end{equation*}
Since the transition density $\Pp$ is a positive function, it is convenient to take only the nonnegative components of $\widetilde{\pmb{\beta}}$ and to replace the negative components by zero. We thus estimate the unknown parameter $\pmb{\beta}^{*}$ by $\widehat{\pmb{\beta}}$ defined by
\begin{equation*}
\widehat{\pmb{\beta}} = \max\{\pmb{0}_{d}; \, \, \widetilde{\pmb{\beta}}\}.
\end{equation*}
Finally, an estimator of the transition density $\Pp$ is defined by:
\begin{equation}\label{eq:wide-P2}
\widehat{\Pp} = \sum_{i = 1}^{d} \widehat{\beta}_{i}\varphi_{i} = \widehat{\pmb{\beta}}^{t} \pmb{\varphi}.
\end{equation}
The estimator $\widehat{\Pp}$ defined in \eqref{eq:wide-P2} is strongly inspired from Sugiyama \& \textit{al.} \cite{sugiyama2010} and the references therein.

In order to study the convergence of $\widehat{\Pp}$, we assume that the density $\Pp$ belongs to a class, $\pmb{\Gg}$ say, of continuous and uniformly bounded functions $\pmb{g}: \RR^{3} \rightarrow \RR$ satisfying, for some constant $M > 0$,
\begin{equation*}
\sup_{\pmb{g} \in \pmb{\Gg}} \Rr(\pmb{g}) \leq M, \quad \text{with} \quad \Rr(\pmb{g}) = \max\left\{\sup_{x \in \RR} \int_{\RR^{2}} \pmb{g}(x,y,z)dydz; \, \, \sup_{x,y,z} \pmb{g}(x,y,z)\right\}.
\end{equation*}
Then \eqref{eq:optim2} can be generalized as follows:
\begin{equation}\label{eq:optim3}
\widehat{\Pp} = \argmin_{\pmb{g} \in \pmb{\Gg}} \left\{ \frac{1}{2|\GG_{n}|} \sum_{u \in \GG_{n}} \int_{\RR^{2}} \pmb{g}(X_{u},y,z)^{2} dydz - \frac{1}{|\GG_{n}|} \sum_{u \in \GG_{n}} \pmb{g}(X_{u},X_{u0},X_{u1}) + \lambda_{n} \Rr(\pmb{g})^{2} \right\}.
\end{equation}
In the sequel, we will study the rate of convergence of $\widehat{\Pp}$. In particular, we will be interested in an upper bound of
\begin{equation*}
\|\widehat{\Pp} - \Pp\|_{L^{2}(\bar{\mu})} = \left(\int (\widehat{\Pp} - \Pp)^{2}(x,y,z) \mu(x)dxdydz\right)^{1/2} \quad \text{where} \quad \bar{\mu}(dx,dy,dz) = \mu(x)dxdydz.
\end{equation*}
\begin{rem}
As we will see, it will be interesting to note that this upper bound is function of the ergodicity rate of the tagged chain $(Y_{m})_{m \in \NN}$. To our best knowledge, this ``new'' phenomenon is not observed for classical processes (for e.g. classical Markov chains). However, we have already highlighted the influence of ergodicity rate of the tagged chain $(Y_{m})_{m \in \NN}$ on the behavior of the additive functionnals of bifurcating Markov chains (see \cite{bitseki23, bitseki24, BD2, BD2020}). 
\end{rem}
Note that \eqref{eq:optim3} implies that
\begin{equation}\label{eq:optim4}
\frac{1}{2}\|\widehat{\Pp} - \Pp\|^{2}_{L^{2}(\bar{\mu})} + \lambda_{n} \,\Rr(\widehat{\Pp})^{2} \leq \left|\int (\widehat{\Pp} - \Pp) d(\mu^{\vartriangle} - \mu^{\vartriangle}_{n})\right| + \frac{1}{2} \left| \int (\widehat{\Pp}^{2} - \Pp^{2}) d(\bar{\mu} - \bar{\mu}_{n}) \right| + \lambda_{n} \, \Rr(\Pp)^{2},
\end{equation}
where the empirical measures $\mu^{\vartriangle}_{n}$ and $\bar{\mu}_{n}$ are defined, for all $\pmb{g} \in \Bb(\RR^{3})$, by 
\begin{equation*}
\int \pmb{g} d\mu^{\vartriangle}_{n} = \frac{1}{|\GG_{n}|} \sum_{u \in \GG_{n}} \pmb{g}(X_{u}^{\vartriangle}) \quad \text{and} \quad \int \pmb{g} d\bar{\mu}_{n} = \frac{1}{|\GG_{n}|} \sum_{u \in \GG_{n}} \int_{\RR^{2}} \pmb{g}(X_{u},y,z) dy dz.
\end{equation*}   
It then follows that to study the rate of convergence of $\widehat{\Pp}$, it suffices to upper bound the right hand side of \eqref{eq:optim4}. This motivated the study of maximal inequality for bifurcating Markov chains doing in Section \ref{sec:max-ineq}. The following Assumption, which completes Assumption \ref{ass:densityq} will be use only for the study of the rate of convergence of our estimator.
\begin{assumption}\label{ass:densityq-b}
Under Assumption \ref{hyp:DenMu}, we assume that
\[\sup_{x,y \, \in \, \RR} \left|\frac{\Qq(x,y)}{\mu(y)}\right| < +\infty \quad \text{and} \quad \sup_{x,y,z \, \in \, \RR} \{\widehat{\Pp}(x,y,z)\} < +\infty.\]
\end{assumption}

\section{Maximal inequalities for bifurcating Markov chains}\label{sec:max-ineq}

In order to study the rate of convergence of the estimator $\widehat{\Pp}$, we develop maximal inequalities for BMCs under bracketing and geometric uniform ergodicity assumptions. In the sequel, $\Gg$ (resp. $\pmb{\Gg}$) will denote the class of continuous functions uniformly bounded defined in $\RR$ (resp. in $\RR^{3}$). For all $g \in \Gg$ (resp. all $\pmb{g} \in \pmb{\Gg}$), we set $\|g\|_{\infty} = \sup_{x \in \RR} |g(x)|$ (resp. $\|\pmb{g}\|_{\infty} = \sup_{x,y,z \in \RR} |\pmb{g}(x,y,z)|$). We will work with the following assumption.
\begin{assumption}\label{ass:geo-erg} The transition $\Qq$ admits a unique invariant probability measure $\mu$
and there exist $R > 0$ and $0 < \alpha < 1$ such that
\begin{equation*}\label{eq:lyapunov}
\big|\Qq^{m}g(x) - \langle \mu,g \rangle\big| \leq R \| g \|_{\infty} \, \alpha^{m} , \quad m \geq 0,
\end{equation*}
for every bounded function $g$ on $\RR$. 
\end{assumption}
\begin{rem}\label{rem:geo-erg}
For the mother-daughters triangle process $X^{\vartriangle}$, Assumption \ref{ass:geo-erg} and \eqref{eq:Qtrin} imply that for all bounded function $\pmb{g}$ defined on $\RR^{3}$, we also have
\begin{equation*}
|(\Qq^{\vt})^{m} \pmb{g} - \langle \mu^{\vartriangle},\pmb{g} \rangle| \leq  R \, \|\pmb{g}\|_{\infty} \, \alpha^{m}  \quad \text{for all $m \in \NN$}.
\end{equation*} 
\end{rem}
For all bounded function $g$ defined on $\RR$ (resp. $\pmb{g}$ defined on $\RR^{3}$), we will set $\widetilde{g} = g - \langle \mu, g\rangle$ (resp. $\widetilde{\pmb{g}} = \pmb{g} - \langle \mu^{\vartriangle}, \pmb{g} \rangle$).

We suppose that Assumptions \ref{hyp:DenMu}, \ref{ass:densityq} and \ref{ass:geo-erg} hold. For every bounded function $g$ defined on $\RR$, we set:
\begin{align}
&c_{\alpha} = \frac{4(1 \vee R^{2})}{2\alpha^{2}(1 - 2\alpha^{2})} \ind_{\{2\alpha < \sqrt{2}\}} + 16(1 \vee R^{2}) \ind_{\{2 \alpha = \sqrt{2}\}} + \frac{8(1 \vee R^{2})}{(2\alpha^{2} - 1)} \ind_{\{2\alpha > \sqrt{2}\}}; \nonumber \\
&c_{1}(g) =  c_{\alpha} \|\Qq g^{2}\|_{\infty} \left(\ind_{2\alpha < \sqrt{2}} + n \ind_{\{2 \alpha = \sqrt{2}\}} + (2\alpha^{2})^{n} \ind_{\{2\alpha > \sqrt{2}\}}\right); \label{eq:c1g}\\
&c_{2}(g) = \tfrac{4}{3} (1 + R \alpha) \|\widetilde{g} \|_{\infty} \ind_{\{2 \alpha \leq 1\}} + \tfrac{2}{3} \alpha^{-1} R (1 + \alpha) \|\Qq g\|_{\infty} \ind_{\{2\alpha > 1\}}. \nonumber
\end{align}
Furthermore, for every bounded function $\pmb{g}$ defined on $\RR^{3}$, we set:
\begin{align}
&c_{1}(\pmb{g}) =  c_{\alpha} \|\Qq \Pp \pmb{g}^{2}\|_{\infty} \left(\ind_{2\alpha < \sqrt{2}} + n \ind_{\{2 \alpha = \sqrt{2}\}} + (2\alpha^{2})^{n} \ind_{\{2\alpha > \sqrt{2}\}}\right);  \label{eq:c1g-bold}\\
&c_{2}(\pmb{g}) = \tfrac{4}{3} (1 + R \alpha) \|\widetilde{\pmb{g}}\|_{\infty} \ind_{\{2 \alpha \leq 1\}} + \tfrac{2}{3} \alpha^{-1} R (1 + \alpha) \|\Qq \Pp \pmb{g}\|_{\infty} \ind_{\{2\alpha > 1\}}. \nonumber
\end{align}

%
%
We also set
\begin{equation*}
v_{n}(\alpha) = |\GG_{n}|^{-1/2} \ind_{\{0 < 2\alpha \leq 1\}} + (2\alpha^{2})^{n/2} \ind_{\{1 < 2\alpha < 2\}}.
\end{equation*}
Then, we have the following concentration inequalities.
\begin{lem}\label{lem:bernstein}
Under Assumptions \ref{hyp:DenMu}, \ref{ass:densityq} and \ref{ass:geo-erg}, we have, for all bounded function $g$ on $\RR$,
\begin{equation*}
\PP\left(|\GG_{n}|^{-1/2} |M_{\GG_{n}}(\widetilde{g})| \geq \delta\right) \leq C \exp\left(- \frac{\delta^{2}}{2(c_{2}(g) \, v_{n}(\alpha) \, \delta \, + \, c_{1}(g))}\right).
\end{equation*}
Moreover, for all bounded function $\pmb{g}$ on $\RR^{3}$, we have
\begin{equation*}
\PP\left(|\GG_{n}|^{-1/2} |M_{\GG_{n}}(\widetilde{\pmb{g}})| \geq \delta\right) \leq C \exp\left(- \frac{\delta^{2}}{2(c_{2}(\pmb{g}) \, v_{n}(\alpha) \, \delta \, + \, c_{1}(\pmb{g}))}\right).
\end{equation*} 
\end{lem}

\begin{rem}
Lemma \ref{lem:bernstein} generalizes the deviation inequalities for empirical means over $\GG_{n}$ obtained in \cite{BHO}. Indeed here, the deviation inequality is valid for all $\alpha \in (0,1)$ which is not the case in \cite{BHO}. However, we omit the proof of this Lemma since it is in the same spirit that the proof of Theorem 4 in \cite{BHO}. Note however that the second inequality can be seen as consequence of Remarks \ref{rem:modau-trig} and \ref{rem:geo-erg}.
\end{rem} 


In view of Lemma \ref{lem:bernstein}, we need to measure the closeness of two functions of $\Gg$ with the help of, say, the $\Qq\text{-}$norm $d_{\Qq}$ defined for all $g \in \Gg$ by: $d_{\Qq}(g) = \sqrt{c_{\alpha} \|\Qq g^{2}\|_{\infty}}.$ For two functions $f$ and $h$ of $\Gg$, the ``bracket'' $[f,h]$ between $f$ and $g$ is defined as:
\begin{equation*}
[f,g] = \{g \in \Gg: \forall x \in \RR, f(x) \leq g(x) \leq h(x)\}.
\end{equation*}
For $\delta > 0$, we denote by $\Nn_{B}(\delta, \Gg, d_{\Qq})$ the minimal number $N$ of brackets $\{[g_{j}^{L},g_{j}^{U}]\}_{j \in \{1,\ldots,N\}}$ such that $d_{\Qq}(g_{j}^{U} - g_{j}^{L}) \leq \delta$ and for all $g \in \Gg$, there is a $j^{*} = j(g) \in \{1, \ldots, N\}$ such that $g \in [g_{j^{*}}^{L};g_{j^{*}}^{U}].$ We take $\Nn(\delta, \Gg, d_{\Qq}) = \infty$ if no finite set of such brackets exists.
\begin{defi}
For all $\delta >0$, the number $\Hh_{B}(\delta, \Gg,d_{\Qq}) = \log(\Nn_{B}(\delta,\Gg,d_{\Qq}))$ is called $\delta\text{-}$entropy with bracketing of $\Gg$ with respect to $d_{\Qq}.$ 
\end{defi}

We set: 
\begin{equation}\label{eq:R-K-kappa}
R_{\sigma} = \sup_{g \in \Gg} \sqrt{c_{1}(g)}, \quad K = \sup_{g \in \Gg} c_{2}(g) \quad \text{and} \quad  \kappa_{n}(\alpha) = \ind_{\{2 \alpha \leq 1\}} + (2\alpha)^{n/2} \ind_{\{1 < 2\alpha < 2\}}.
\end{equation}
\begin{thm}\label{thm:max-ineq}
Let $C$, $C_{0}$ and $C_{1}$ be positive constants and let $(a_{n}, n \in \NN)$ be a sequence of real numbers such that:
\begin{align}
& K a_{n} \leq C_{1} \sqrt{|\GG_{n}|}R_{\sigma}^{2};   
\label{eq:cond-an1}\\  
&a_{n} \leq 8 \sqrt{|\GG_{n}|} R_{\sigma}; \label{eq:cond-an2}\\
&a_{n}  > C_{0} \, \kappa_{n}(\alpha) \, \left(\int_{\tfrac{a_{n}}{2^{6}\sqrt{|\GG_{n}|}}}^{R_{\sigma}} \Hh_{B}^{1/2}(x, \Gg, d_{\Qq})dx \vee R_{\sigma}\right); \label{eq:cond-an3} \\
&C_{0}^{2} \geq C^{2}(C_{1} + 1). \label{eq:cond-C0-C}
\end{align}
Under Assumptions \ref{hyp:DenMu}, \ref{ass:densityq} and \ref{ass:geo-erg}, there exist constants $c_{1}$ and $c_{2}$, depending on $C$, $C_{0}$ and $C_{1}$, such that
\begin{equation*}
\PP\left( \sup_{g \in \Gg} \left| \frac{1}{\sqrt{|\GG_{n}|}} \sum_{u \in \GG_{n}} (g - \langle \mu,g \rangle)(X_{u}) \right| > a_{n} \right) \leq c_{1} \, \exp\left( - \frac{c_{2} \, a_{n}^{2}}{\kappa_{n}(\alpha)^{2} \, R_{\sigma}^{2}} \right).
\end{equation*}
\end{thm}
Using Remarks \ref{rem:modau-trig} and \ref{rem:geo-erg} and the second inequality of Lemma \ref{lem:bernstein}, Theorem \ref{thm:max-ineq} can be extended to the mother-daughters triangle process as follows. We set: 
\begin{equation*} 
\pmb{R_{\sigma}} = \sup_{\pmb{g} \in \pmb{\Gg}} \sqrt{c_{\alpha} \|\Qq \Pp \pmb{g}^{2}\|_{\infty}}; \quad \pmb{K} = \sup_{\pmb{g} \in \pmb{\Gg}} c_{2}(\pmb{g}); \quad d_{\Qq^{\vartriangle}}(\pmb{g}) = \sqrt{c_{\alpha} \|\Qq\Pp \pmb{g}^{2}\|_{\infty}}.
\end{equation*}
Mimicking the case of $\Gg$, the closeness of two functions of $\pmb{\Gg}$ is measured with the help of the $\Qq^{\vartriangle}\text{-}$norm $d_{\Qq^{\vartriangle}}.$
\begin{cor}\label{cor:max-ineq-T}
Under the settings of Theorem \ref{thm:max-ineq} by replacing: $R_{\sigma}$ by $\pmb{R_{\sigma}}$, $K$ by $\pmb{K}$, $d_{\Qq}$ by $d_{\Qq^{\vartriangle}}$ and $\Gg$ by $\pmb{\Gg}$, we have
\begin{equation*}
\PP\left( \sup_{\pmb{g} \in \pmb{\Gg}} \left| \frac{1}{\sqrt{|\GG_{n}|}} \sum_{u \in \GG_{n}} (\pmb{g} - \langle \mu^{\vartriangle}, \pmb{g} \rangle)(X^{\vartriangle}_{u}) \right| > a_{n} \right) \leq c_{1} \exp\left( - \frac{c_{2} \, a_{n}^{2}}{\kappa_{n}(\alpha)^{2} \, \pmb{R_{\sigma}}^{2}} \right).
\end{equation*}
\end{cor}

\begin{rem}
We stress that unlike the classical cases (see for e.g. \cite{van2000applications}), the previous maximal inequalities depend on the geometric rate of convergence $\alpha$ defined in Assumption \ref{ass:geo-erg}. We also stress that unlike the central limit theorem studied in \cite{BD2, BD2020}, we see two main regimes appear: $0 < 2\alpha \leq 1$ and $1 < 2\alpha < 2$. As explained in \cite{bitseki23}, this is due to the fact that for large deviations, there is a phase transition at $\alpha = 1/2.$
 \end{rem}

Now, let $g_{0} \in \Gg$
and $\pmb{g}_{0} \in \pmb{\Gg}.$ 
We are interested in the behaviour of 
\begin{equation*}
\sum_{u \in \GG_{n}} (\widetilde{g} - \widetilde{g}_{0})(X_{u}) \quad \forall g \in \Gg \quad
\text{and} \quad \sum_{u \in \GG_{n}} (\widetilde{\pmb{g}} - \widetilde{\pmb{g}}_{0})(X_{u}^{\vartriangle}) \quad \forall \pmb{g} \in \pmb{\Gg}.
\end{equation*}
For that purpose, we do the following assumption.
\begin{assumption}\label{ass:modulus-c}
The classes of functions $\Gg$ and $\pmb{\Gg}$ satisfy
\begin{multline}\label{eq:modulus-c}
\max\left\{\sup_{g \in \Gg} \|g - g_{0}\|_{\infty}, \, \, \sup_{\pmb{g} \in \pmb{\Gg}} \|\pmb{g} - \pmb{g_{0}}\|_{\infty}\right\} < 1, \quad \text{and} \\ 
\max\Big\{ \Hh_{B}(\delta,\Gg,d_{\Qq}), \Hh_{B}(\delta,\pmb{\Gg},d_{\Qq^{\vartriangle}})\Big\} \leq A \delta^{-\gamma}
\end{multline}
for all $\delta > 0$, where $A$ is a positive constant. 
\end{assumption}
\begin{rem}
Note that the second inequality in \eqref{eq:modulus-c} implies in particular that
\begin{equation*}
\max\left\{\int_{0}^{\delta} \Hh_{B}^{1/2}(x,\Gg,d_{\Qq})dx, \int_{0}^{\delta} \Hh_{B}^{1/2}(x,\pmb{\Gg},d_{\Qq^{\vartriangle}})dx\right\} \leq \frac{A^{1/2}}{1 - \frac{\gamma}{2}} \, \delta^{1-\gamma/2} = A_{0} \, \delta^{1-\gamma/2} \quad \forall \delta > 0,
\end{equation*}
We emphasise that here, the complexity parameter $\gamma$ is implicitly a function of the geometric ergodicity rate $\alpha$ through the condition \eqref{eq:cond-an3}. To our best knowledge, this phenomenon is not observed for the classical random processes.  
\end{rem}

\begin{rem}
The following choices of $g_{0}$ and $\pmb{g}_{0}$ will be relevant in the study of the rate of convergence of the least square estimator $\widehat{\Pp}$:
\begin{equation*}
g_{0}(x) = \int_{\RR^{2}} \Pp^{2}(x,y,z) dydz \quad \text{and} \quad \pmb{g}_{0}(x,x_{0},x_{1}) = \Pp(x,x_{0},x_{1}) \quad \forall x,x_{0},x_{1} \in \RR.
\end{equation*} 
\end{rem}

For all $\delta > 0$, we set $\Gg(\delta) = \{g \in \Gg: d_{\Qq}(g,g_{0}) \leq \delta\}$ and $\pmb{\Gg}(\delta) = \{\pmb{g} \in \pmb{\Gg}: d_{\Qq^{\vartriangle}}(\pmb{g},\pmb{g}_{0}) \leq \delta\}.$
We do the following assumption.
\begin{assumption}\label{ass:c2g-g0}
For all $\delta \in (0,1)$, 
\begin{equation*}
\sup_{g \in \Gg(\delta)} c_{2}(g-g_{0}) \leq K \quad \text{and} \quad \sup_{\pmb{g} \in \pmb{\Gg}(\delta)} c_{2}(\pmb{g} - \pmb{g}_{0}) \leq \pmb{K}
\end{equation*}
\end{assumption}

Recall the parameter $\kappa_{n}(\alpha)$ defined in \eqref{eq:R-K-kappa}. We set 
\begin{equation}\label{eq:delta-n}
\mathfrak{e}(\alpha) = \ind_{\{0 < 2\alpha \leq 1\}} + \frac{\log(1/\alpha)}{\log(2)} \ind_{\{1 < 2\alpha < 2\}} \quad \text{and} \quad \delta_{n} = |\GG_{n}|^{- \mathfrak{e}(\alpha)/(2+\gamma)}. 
\end{equation}
The following Lemmas, which are stated only for $\Gg$, are the consequence of Theorem \ref{thm:max-ineq}.  The results and the proofs are the same for $\pmb{\Gg}$ and are therefore omitted (replace only $\Gg$ by $\pmb{\Gg}$ and $d_{\Qq}$ by $d_{\Qq^{\vartriangle}}$ in Lemmas \ref{lem:modulus-c} and \ref{lem:modulus-c2}). We will also omitted the proof of Lemma \ref{lem:modulus-c2} since it is an extension of Lemma \ref{lem:modulus-c}.

\begin{lem}\label{lem:modulus-c}
Under Assumptions of Theorem \ref{thm:max-ineq} and Assumptions \ref{ass:modulus-c} and \ref{ass:c2g-g0}, we have for all $\alpha \in (0,1)$, for some constants $C$ and $T$ depending on $\gamma$ and $A$, and for $n$ large enough:  
\begin{equation}\label{eq:modulus-c1}
\PP\left( \sup_{g \in \Gg(\delta_{n})} \left| \frac{1}{|\GG_{n}|} \sum_{u \in \GG_{n}} (\widetilde{g} - \widetilde{g}_{0})(X_{u}) \right| > T  \, \delta_{n}^{2} \right) \leq C \exp\left(- C \, |\GG_{n}|^{\mathfrak{e}(\alpha)\gamma/(2 + \gamma)} \right);  
\end{equation}
\begin{equation}\label{eq:modulus-c2}
\PP\left( \sup_{g \in \Gg; \, d_{\Qq}(g,g_{0}) \, >  \, \delta_{n}} \frac{\left| \frac{1}{\sqrt{|\GG_{n}|}} \sum_{u \in \GG_{n}} (\widetilde{g} - \widetilde{g}_{0})(X_{u}) \right|}{d_{\Qq}(g,g_{0})^{1 - \gamma/2} } \geq \kappa_{n}(\alpha) T\right) \leq C \exp\left( - C \,T \right).
\end{equation}
\end{lem}

Moreover, Lemma  \ref{lem:modulus-c} can be extended as follows. Let $I : \Gg \mapsto [1, \infty)$ be a map such that for some $M > 0$ and for all $g \in \Gg,$ $I(g) \leq M.$ Then we have the following.

\begin{lem}\label{lem:modulus-c2}
Under Assumptions of Theorem \ref{thm:max-ineq} and Assumptions \ref{ass:modulus-c} and \ref{ass:c2g-g0}, we have for all $\alpha \in (0,1)$, for some constants $C$ and $T$ depending on $\gamma$ and $A$, and for $n$ large enough:   
\begin{equation*}
\PP\left( \sup_{g \in \Gg; \, d_{\Qq}(g,g_{0}) \, \leq \, \delta_{n} \, I(g)} \left| \frac{1}{|\GG_{n}|} \sum_{u \in \GG_{n}} (\widetilde{g} - \widetilde{g}_{0})(X_{u}) \right| > T  \, \delta_{n}^{2} \right) \leq C \exp\left(- C \, |\GG_{n}|^{\mathfrak{e}(\alpha)\gamma/(2 + \gamma)} \right);  
\end{equation*}
\begin{equation*}
\PP\left( \sup_{g \in \Gg; \, d_{\Qq}(g,g_{0}) \, > \, \delta_{n} \, I(g)} \frac{\left| \frac{1}{\sqrt{|\GG_{n}|}} \sum_{u \in \GG_{n}} (\widetilde{g} - \widetilde{g}_{0})(X_{u}) \right|}{d_{\Qq}(g,g_{0})^{1 - \gamma/2} \, I(g)^{\gamma/2}} \geq \kappa_{n}(\alpha) \, T\right) \leq C \exp\left( - C \, T \right).
\end{equation*}
\end{lem}
 
\section{Study of the convergence}

We are now able to study the convergence rate of $\widehat{\Pp}$. For the definition of stochastic order symbols $\bigO_{\PP}$ and $\smallO_{\PP}$, we refer to \cite{van2000applications}, Section 2.1. Recall $\kappa_{n}(\alpha)$ defined in \eqref{eq:R-K-kappa} and $\delta_{n}$ defined in \eqref{eq:delta-n}.
\begin{thm}\label{thm:cv-rate}
Under the Assumptions of Lemma \ref{lem:modulus-c2} and Assumption \ref{ass:densityq-b}, if $\lambda_{n} \longrightarrow 0$ and $\lambda_{n}^{-1} = \smallO_{\PP}\left(\kappa_{n}(\alpha)^{-4/(2+\gamma)} \delta_{n}^{-2}\right),$ then
\begin{equation*}
\|\widehat{\Pp} - \Pp\|_{L^{2}(\bar{\mu})} = \bigO_{\PP}(\lambda_{n}^{1/2}).
\end{equation*}
\end{thm}

\begin{rem}
One can observe that the convergence rate of $\widehat{\Pp}$ is a function of the geometric ergodic rate $\alpha$. Indeed, according to the value of the geometric ergodic rate $\alpha \in (0,1)$, one can observe that the convergence rate of $\widehat{\Pp}$ is constant when $0 < 2\alpha \leq 1$ and it deteriorates with the value of $\alpha$ when $2\alpha \in (1,2)$. Here the phase transition occurs at point $\alpha = 1/2.$ This differs from \cite{BD2, BD2020} where phase transition for the fluctuations is observed  at point $\alpha = 1/\sqrt{2}.$ Nevertheless, this type of behavior, that is a phase transition at point $\alpha = 1/2$, has already been observed in \cite{bitseki23}.
\end{rem}

\section{Numerical study}
\subsection{Estimating the density transition of the NBAR process}
We consider the NBAR process $(X_{u}, u \in \TT)$ given in \cite{BO2018, BO17}. It is defined as follows. Let $f_{0}$ and $f_{1}$ be two real functions defined on $\RR$. Let $\nu$ be a probability measure on $(\RR, \Bb(\RR))$. Then, $X_{\emptyset}$ is distributed as $\nu$ and for all $u \in \TT$, we have
$$
X_{u0} = f_{0}(X_{u}) + \ep_{u0} \quad \text{and} \quad X_{u1} = f_{1}(X_{u}) + \ep_{u1},
$$
where $((\ep_{u0}, \ep_{u1}), u \in \TT)$ is a sequence of independent and bivariate random variables with common density, $g_{\ep}$ say. Then, $(X_{u}, u \in \TT)$ is a bifurcating Markov chain with
$$
\Pp(x,dy,dz) = g_{\ep}(y - f_{0}(x), z - f_{1}(x))dydz \quad \text{and} \quad Q(x,dy) = \frac{1}{2}\Big(g_{0}\big(y-f_0(x)) + g_{1}\big(y-f_1(x)\big) \Big)dy,
$$
where the marginal densities $g_{0}$ and $g_{1}$ are defined by
$$
g_0(\cdot) = \int_\RR g_\ep(\cdot,y)dy \quad \text{and} \quad g_1(\cdot) = \int_\RR g_\ep(x,\cdot)dx.
$$
For our numerical study, we choose the functions $f_{0}$, $f_{1}$ and $g_{\ep}$ as follows: for all $x,y \in \RR$, 
$$
f_{0}(x) = f_{1}(x) = \sinc(2\pi x) \quad  \text{and} \quad g_{\ep}(x,y) = \frac{1}{2\sigma^{2}\pi\sqrt{1 - \rho^{2}}} \exp(\frac{x^{2} - 2\rho xy + y^{2}}{2\sigma^{2}(1-\rho^{2})}).
$$
We will take $\sigma = 1$ and $\rho = 0.3$. Our aim is to reconstruct the transition density $\Pp$ given by
$$
\Pp(x,y,z) = g_{\ep}(y - f_{0}(x), z - f_{1}(x)).
$$
We simulate a sample $(X^{\vt}_{u}, u \in \GG_{n})$ of NBAR process with $n = 16$, $f_{0}$, $f_{1}$ and $g_{\ep}$ given below. Let us note that with these choices of $f_{0}$, $f_{1}$ and $g_{\ep}$, Assumptions \ref{hyp:DenMu},\ref{ass:densityq}, \ref{ass:geo-erg}  are satisfied. 

We consider the gaussian basis of functions $\pmb{\varphi} = (\varphi_{1}, \ldots, \varphi_{d})^{t},$ where $d = \min(1000, 2^{n})$ and for all $j \in \{1, \ldots, d\}$, 
$$
\varphi_{j}(x,y,z) = \exp\left(-\frac{(x-c_{j})^{2}}{2 \tau^{2}}\right) \exp\left(-\frac{(y-c_{j0})^{2}}{2 \tau^{2}}\right) \exp\left(-\frac{(z-c_{j1})^{2}}{2 \tau^{2}}\right), 
$$
where the real parameter $\tau > 0$ must be chosen carefully. The centers $c_{j}, c_{j0}, c_{j1}$ are chosen in the sample $(X^{\vt}_{u}, u \in \GG_{n})$ as follows. We randomly chose $c_{j}$ in $(X_{u}, u \in \GG_{n})$ and if $c_{j} = X_{u}$, $u \in \GG_{n}$, we set $c_{j0} = X_{u0}$ and $c_{j1} = X_{u1}$. With this choice, $\overline{\varphi}_{i,j}$ defined in \eqref{eq:basis} is given by
$$
\overline{\varphi}_{i,j}(x) = \pi \tau^{2} \exp\left( - \frac{2(x-c_{j})^{2} + (c_{j0} - c_{i0})^{2} + 2(x - c_{i})^{2} + (c_{j1} - c_{i1})^{2}}{4 \tau^{2}} \right).
$$
Now, in the test phase, we have to renormalize the estimator given in \eqref{eq:wide-P2} in order to assure that, for a fixed point $x$, $\widehat{\Pp}(x,\cdot,\cdot)$ is a density function. The final estimator obtained with the gaussian basis $\pmb{\varphi}$ is then
$$
\widehat{\Pp}(x,y,z) = \frac{\sum_{i=1}^{d} \widehat{\beta}_{i} \varphi_{i}(x,y,z)}{2\pi \tau^{2} \sum_{j=1}^{d} \widehat{\beta}_{j} \exp\left( - \frac{(x-c_{j})^{2}}{2\tau^{2}}\right)}.
$$
\subsection{Selection of the parameters by Cross-validation method}
We choose the regularization parameters $\lambda$ and the variance parameters $\tau$ which minimize the integrated squared error $ISE(\widehat{\Pp}$ or the Kullback-Leibler error $KL(\widehat{\Pp})$ defined respectively by
\begin{multline*}
ISE(\widehat{\Pp}) = \frac{1}{2}\int_{\RR^{3}} \left( \widehat{\Pp}(x,y,z) - \Pp(x,y,z) \right)^{2} \mu(x) dxdydz, \\ KL(\widehat{\Pp}) = \int_{\RR^{3}} \mu(x) \Pp(x,y,z) \log\left( \frac{\Pp(x,y,z)}{\widehat{\Pp}(x,y,z)} \right) dxdydz,
\end{multline*}
where $\mu(\cdot)$ is the invariant density associated to the NBAR process. Now, minimizing $ISE(\widehat{\Pp})$ or $KL(\widehat{\Pp})$ is equivalent to minimize the functions $S$ or $K$ defined respectively by
\begin{multline*}
S(\lambda,\tau) = \int_{\RR^{3}} \widehat{\Pp}(x,y,z)^{2} \mu(x) dx - 2 \int_{\RR^{3}} \widehat{\Pp}(x,y,z) \Pp(x,y,z) \mu(x) dxdydz \\
K(\lambda,\tau) = - \int_{\RR^{3}} \mu(x) \Pp(x,y,z) \log\left( \widehat{\Pp}(x,y,z) \right)dxdydz.
\end{multline*}
The method to select the parameters is the following.
\begin{enumerate}
\item We divide the sample $(X^{\vt}_{u})_{u \in \GG_{n}}$ into $K$ disjoints subsamples $\{(X^{\vt}_{u})_{u \in \GG_{n}^{(k)}}, k \in \{1, \ldots, K\}\},$ with $(\GG_{n}^{(k)}, k \in \{1, \ldots, K\})$ a partition of $\GG_{n}.$
\item For each subsample $(X_{u})_{u \in \GG_{n}^{(k)}}$: 
\begin{enumerate}
\item We set $\widehat{\Pp}_{\GG_{n}[-k]}$ the estimator of $\Pp$ obtaining using the subsample  $(X^{\vt}_{u})_{u \in \GG_{n}} \mysetminus (X^{\vt}_{u})_{u \in \GG_{n}^{(k)}},$ where for two sets $A$ and $B$, $B \mysetminus A$ denotes the set of elements in $B$ but not in $A.$
\item Next, we approximate the integrated squared and the Kullback-Leibler errors for the estimator $\widehat{\Pp}_{\GG_{n}[-k]}$ by
\begin{align*}
&\widehat{S}_{\GG_{n}^{(k)}} = \frac{1}{2|\GG_{n}^{(k)}|} \sum_{u \in \GG_{n}^{(k)}} \int_{\RR^{2}} \left(\widehat{\Pp}_{\GG_{n}[-k]}(X_{u},y,z)\right)^{2}dydz - \frac{1}{|\GG_{n}^{(k)}|} \sum_{u \in \GG_{n}^{(k)}} \widehat{\Pp}_{\GG_{n}[-k]}(X_{u}, X_{u0}, X_{u1});\\
&\widehat{K}_{\GG_{n}^{(k)}} = - \frac{1}{|\GG_{n}^{(k)}|} \sum_{u \in \GG_{n}^{(k)}} \log\left( \widehat{\Pp}_{\GG_{n}[-k]}(X_{u}, X_{u0}, X_{u1}) \right).
\end{align*}
\end{enumerate}
\item Let $\Ll = \{\lambda_{1}, \ldots, \lambda_{m_{1}}\}$ and $\Tt = \{\tau_{1}, \ldots, \tau_{m_{2}}\}$ be a grid of regularization and variance parameters. Then, we select the couple $(\lambda, \tau) \in \Ll \times \Tt$ which minimize one of the sums
$$
\frac{1}{K}\sum_{k=1}^{K} \widehat{S}_{\GG_{n}^{(k)}} \quad \text{or} \quad \frac{1}{K}\sum_{k=1}^{K} \widehat{K}_{\GG_{n}^{(k)}}.
$$
\end{enumerate}
For our numerical study, we will take $\Tt = \{0.0316,    0.075,    0.1778,    0.4217,    1,   2.3714,    5.6234,   13.3352,   31.6228\}$ and $\Ll = \{0.001,    0.0032,    0.01,   0.0316,    0.1,    0.3162,    1,    3.1623,   10\}
\}$.
\begin{rem}
As we have already mentioned, the motivation for considering the estimators $\widehat{S}_{\GG_{n}^{(k)}} $ and $\widehat{K}_{\GG_{n}^{(k)}}$ come from Guyon's work \cite{Guyon}. However, theoretical studies of these estimators are needed and are left for future works. We stress that we have deliberately chosen to fix an arbitrary value for the dimension $d$, although it is possible to choose it by cross validation. 
\end{rem}
 \begin{figure}[!ht]
\includegraphics[scale=.60]{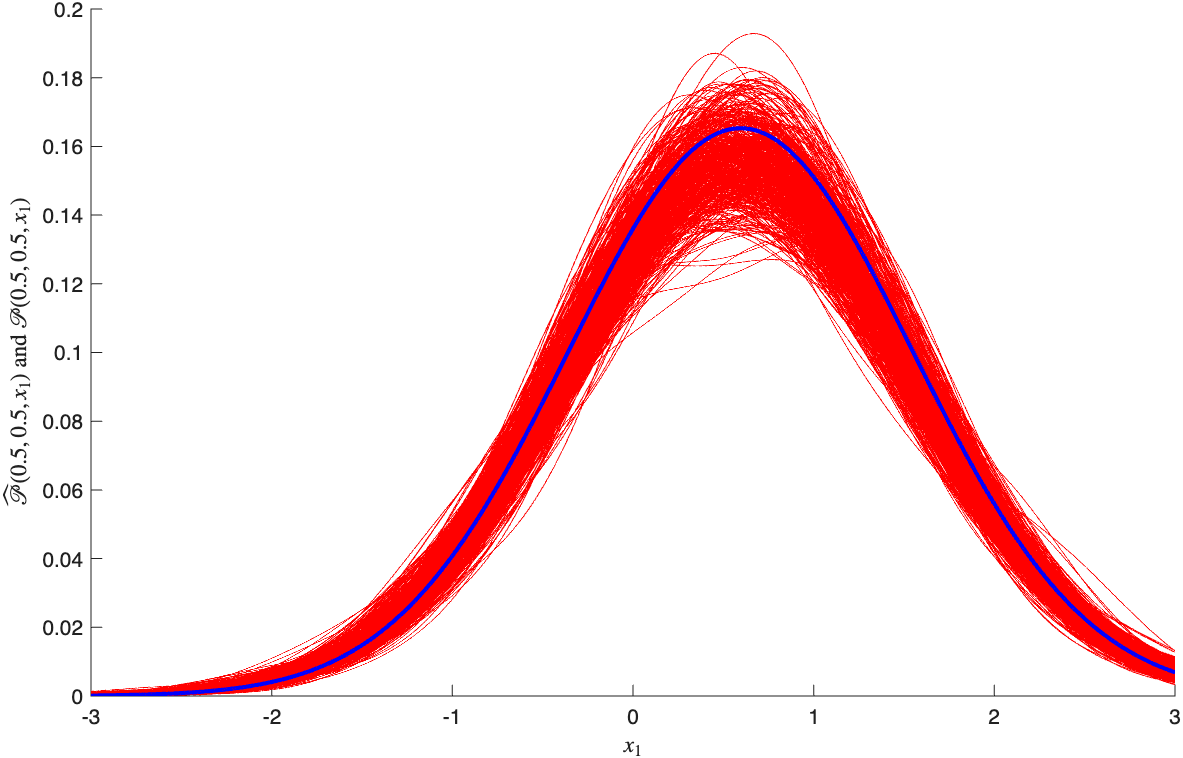}
\centering
\caption{Plot of 500 estimators $\widehat{\Pp}(0.5,0.5,\cdot)$ on the interval $[-3,3]$ and generation $\GG_{13}$ (red solid lines) from independent copies of $X$. The blue solid line represents the function $\Pp(0.5,0.5,\cdot)$ to estimate.}
\label{figtree}
\end{figure}
\begin{figure}[!ht]
\includegraphics[scale=.60]{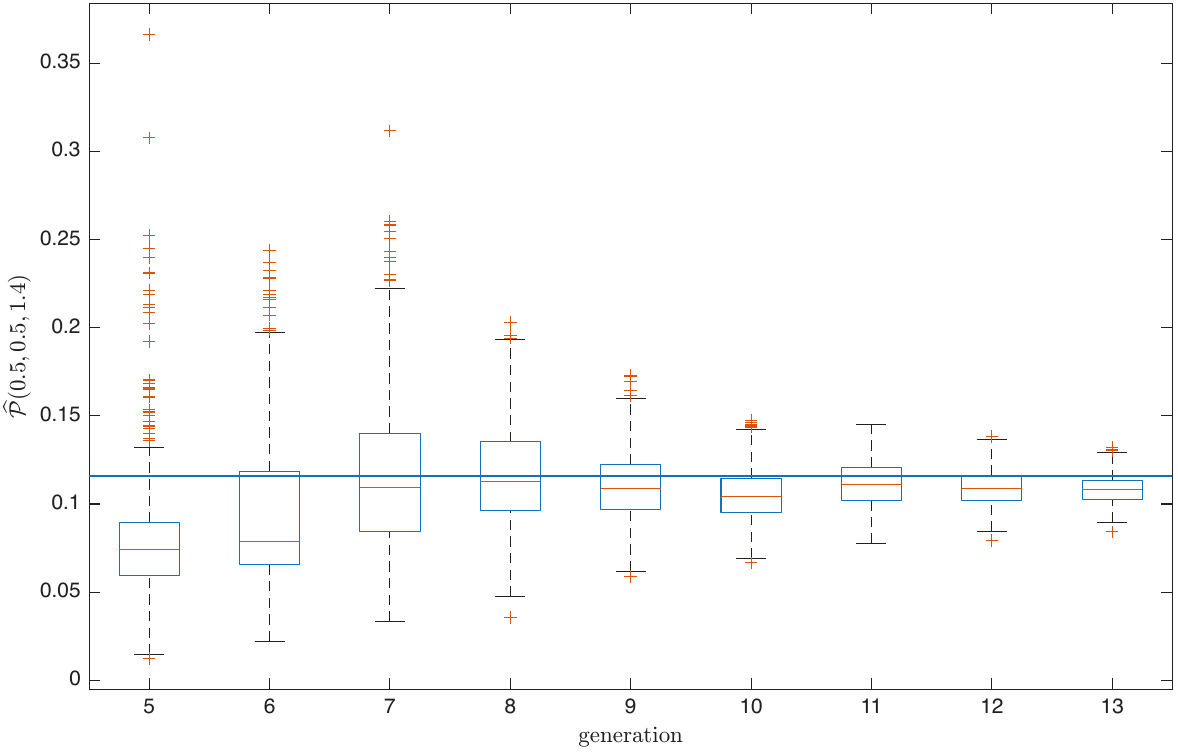}
\centering
\caption{Boxplots of the estimation $\widehat{\Pp}(0.5,0.5,1.4)$ versus the number of generation. The solid line in blue is the true value $\Pp(0.5,0.5,1.4)$.}
\label{figtree}
\end{figure} 

\begin{figure}[!ht]
	\centering
	\begin{subfigure}{0.45\textwidth} 
		\includegraphics[width=\textwidth]{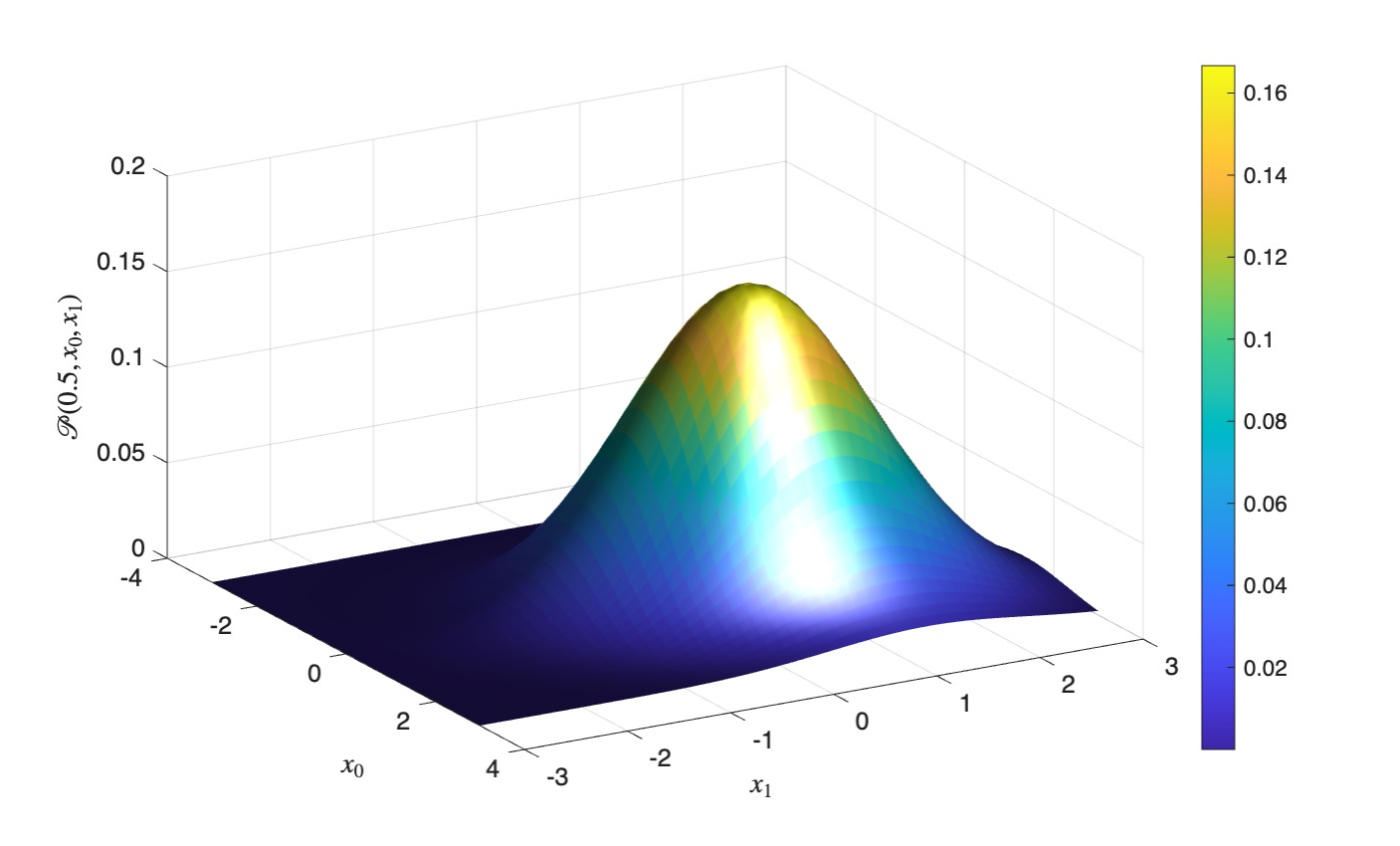}
	\end{subfigure}
	\begin{subfigure}{0.45\textwidth} 
		\includegraphics[width=\textwidth]{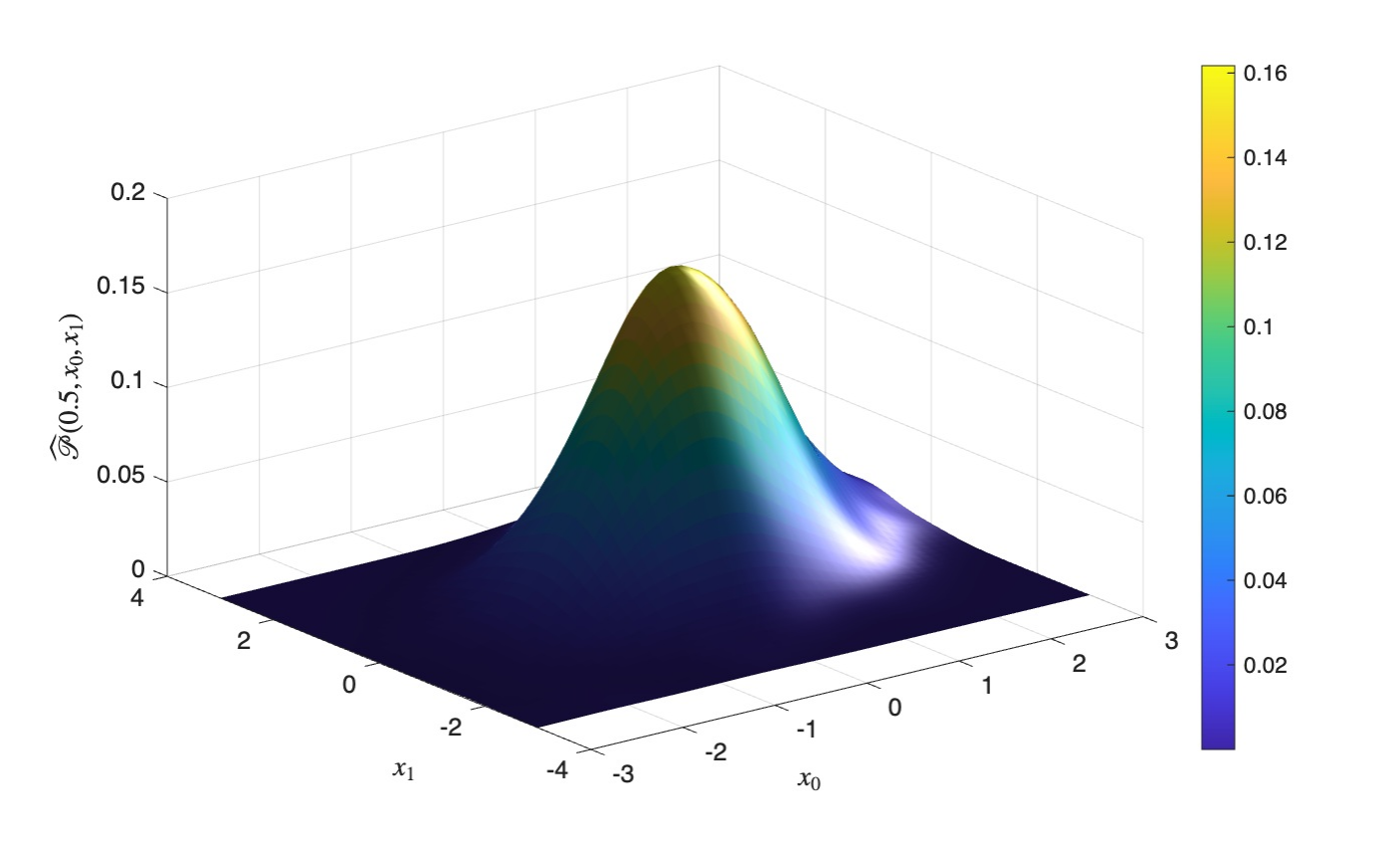}
	\end{subfigure}
	\vspace*{-0.5cm}
        \caption{Plot of the function $\Pp(0.5,\cdot,\cdot)$ (left) and its estimator $\widehat{\Pp}(0.5,\cdot,\cdot)$ (right) on $[-3,3]\times[-3,3]$ and $\GG_{16}$} \label{fig5}
\end{figure}
\begin{figure}[!ht]
\includegraphics[scale=.60]{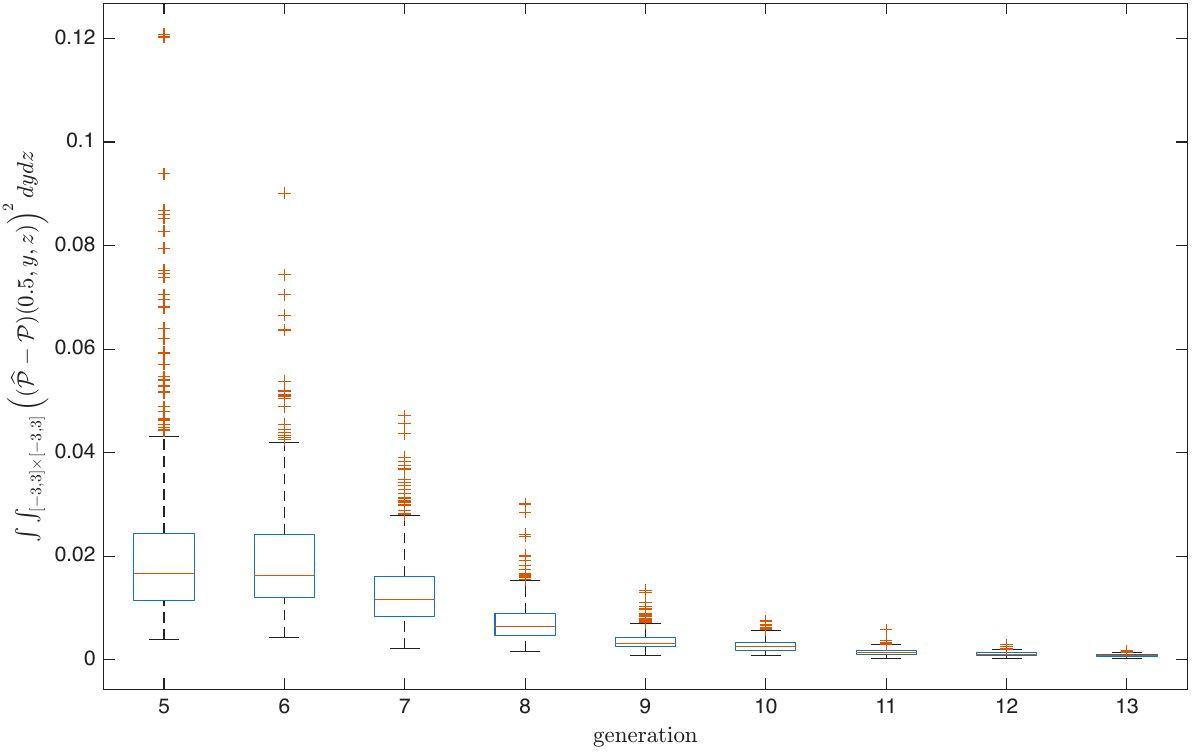}
\centering
\caption{Boxplots of the error $\int\int_{[-3,3]\times[-3,3]}\left((\widehat{\mathcal{P}} - \mathcal{P})(0.5,y,z)\right)^2dydz$ versus the number of generation.}
\label{figtree}
\end{figure} 
\begin{rem}
The comparison of the studied in this paper with the kernel method has been deliberately omitted. It will be covered in detail in a future work.
\end{rem}
\section{Appendix}

\subsection{Proof of Theorem \ref{thm:cv-rate}}
Recalling \eqref{eq:optim4}, we have
\begin{equation*}
\frac{1}{2}\|\widehat{\Pp} - \Pp\|^{2}_{L^{2}(\bar{\mu})} + \lambda_{n} \,\Rr(\widehat{\Pp})^{2} \leq \left|\int (\widehat{\Pp} - \Pp) d(\mu^{\vartriangle} - \mu^{\vartriangle}_{n})\right| + \frac{1}{2} \left| \int (\widehat{\Pp}^{2} - \Pp^{2}) d(\bar{\mu} - \bar{\mu}_{n}) \right| + \lambda_{n} \, \Rr(\Pp)^{2}.
\end{equation*}
We consider the function $p$ and $\widehat{p}$ defined for all $x \in \RR$ by:
\begin{equation*}
p(x) = \int_{\RR^{2}} \Pp^{2}(x,y,z)dydz \quad \text{and} \quad \widehat{p}(x) = \int_{\RR^{2}} \widehat{\Pp}^{2}(x,y,z)dydz.
\end{equation*}
Then we can write
\begin{equation*}
\int (\widehat{\Pp}^{2} - \Pp^{2}) d(\bar{\mu} - \bar{\mu}_{n}) = \int (\widehat{p} - p) d(\mu_{n} - \mu),
\end{equation*}
where the empirical measure $\mu_{n}$ is defined for all $g \in \Gg$ by $\int g d\mu_{n} = |\GG_{n}|^{-1} \sum_{u \in \GG_{n}} g(X_{u}).
$
%
Using Lemma \ref{lem:modulus-c2}, with $p$ and $\Pp$ instead of $g_{0}$ and $\pmb{g}_{0}$, we get
\begin{align}
&\label{eq:BPmuD} \left| \int (\widehat{\Pp}^{2} - \Pp^{2}) d(\bar{\mu} - \bar{\mu}_{n}) \right| \\ 
&\hspace{2cm}\nonumber = \bigO_{\PP}\left( \kappa_{n}(\alpha) |\GG_{n}|^{-1/2} d_{\Qq}(\widehat{p},p)^{1 - \gamma/2} \Rr(\widehat{\Pp})^{\gamma/2}  \vee \kappa_{n}(\alpha) |\GG_{n}|^{-2/(2+\gamma)}\Rr(\widehat{\Pp})\right); \\
&\label{eq:BPmuB} \left|\int (\widehat{\Pp} - \Pp) d(\mu^{\vartriangle} - \mu^{\vartriangle}_{n})\right| \\ 
&\hspace{2cm}\nonumber =  \bigO_{\PP}\left( \kappa_{n}(\alpha) |\GG_{n}|^{-1/2} d_{\Qq^{\vartriangle}}(\widehat{\Pp},\Pp)^{1 - \gamma/2} \Rr(\widehat{\Pp})^{\gamma/2}  \vee \kappa_{n}(\alpha) |\GG_{n}|^{-2/(2+\gamma)}\Rr(\widehat{\Pp})\right). 
\end{align}
Now, for all $t \in \RR$, we have
\begin{align*}
\Qq((\widehat{p} - p)^{2})(t) & = \int_{\RR} \left(\int_{\RR^{2}} \left(\widehat{\Pp}^{2}(x,y,z) - \Pp^{2}(x,y,z)\right)dydz\right)^{2} \Qq(t,x)dx \\
&\leq 2 \int_{\RR^{3}} (\widehat{\Pp} - \Pp)^{2}(x,y,z) (\widehat{\Pp} + \Pp)(x,y,z) \Qq(t,x) dxdydz \\
&\leq C \int_{\RR^{3}} (\widehat{\Pp} - \Pp)^{2}(x,y,z) \mu(x) dx dy dz = C \|\widehat{\Pp} - \Pp\|^{2}_{L^{2}(\bar{\mu})}, 
\end{align*}
where for the first inequality we use $\widehat{\Pp}^{2} - \Pp^{2} = (\widehat{\Pp} - \Pp)(\widehat{\Pp} + \Pp),$ the fact that $\Pp(x,\cdot,\cdot)$ and $\widehat{\Pp}(x,\cdot,\cdot)$ are probability densities and the Cauchy-Schwartz inequality; for the second inequality, we divide and multiply by $\mu(x)$ and we use Assumptions \ref{ass:densityq} and \ref{ass:densityq-b}. The last inequality implies that $\|\Qq((\widehat{p} - p)^{2})\|_{\infty} < C \|\widehat{\Pp} - \Pp\|^{2}_{L^{2}(\bar{\mu})}$ and then that $d_{\Qq}(\widehat{p}, p) \leq C \|\widehat{\Pp} - \Pp\|_{L^{2}(\bar{\mu})}.$ In the same way, we prove that $d_{\Qq^{\vartriangle}}(\widehat{\Pp}, \Pp) \leq C \|\widehat{\Pp} - \Pp\|_{L^{2}(\bar{\mu})}.$ Using these upper bounds of $d_{\Qq}(\widehat{p},p)$ and $d_{\Qq^{\vartriangle}}(\widehat{\Pp},\Pp)$ in \eqref{eq:BPmuD} and \eqref{eq:BPmuB}, it follows from \eqref{eq:optim4} that
\begin{align*}
\frac{1}{2}\|\widehat{\Pp} - \Pp\|^{2}_{L^{2}(\bar{\mu})} + \lambda_{n} \,\Rr(\widehat{\Pp})^{2} &\leq \bigO_{\PP}\left( \kappa_{n}(\alpha) |\GG_{n}|^{-1/2} \|\widehat{\Pp} - \Pp\|_{L^{2}(\bar{\mu})}^{1 - \gamma/2} \Rr(\widehat{\Pp})^{\gamma/2}  \vee  |\GG_{n}|^{-2 \mathfrak{e}(\alpha)/(2+\gamma)}\Rr(\widehat{\Pp}) \right) \\ 
 &  +  \lambda_{n} \, \Rr(\Pp)^{2}.
\end{align*}
Now, we can distinguish three cases:
\begin{align}
&\frac{1}{2}\|\widehat{\Pp} - \Pp\|^{2}_{L^{2}(\bar{\mu})} + \lambda_{n} \,\Rr(\widehat{\Pp})^{2} \leq \bigO_{\PP}(\lambda_{n}) \label{eq:case1};\\
&\frac{1}{2}\|\widehat{\Pp} - \Pp\|^{2}_{L^{2}(\bar{\mu})} + \lambda_{n} \,\Rr(\widehat{\Pp})^{2} \leq \bigO_{\PP}\left(\kappa_{n}(\alpha) |\GG_{n}|^{-1/2} \|\widehat{\Pp} - \Pp\|_{L^{2}(\bar{\mu})}^{1 - \gamma/2} \Rr(\widehat{\Pp})^{\gamma/2}\right); \label{eq:case2}\\
&\frac{1}{2}\|\widehat{\Pp} - \Pp\|^{2}_{L^{2}(\bar{\mu})} + \lambda_{n} \,\Rr(\widehat{\Pp})^{2} \leq \bigO_{\PP}\left(|\GG_{n}|^{-2\mathfrak{e}(\alpha)/(2+\gamma)}\Rr(\widehat{\Pp}) \right). \label{eq:case3}
\end{align}

\noindent {\bf Case \eqref{eq:case2}:} We have
\begin{align}
&\|\widehat{\Pp} - \Pp\|^{2}_{L^{2}(\bar{\mu})} \leq \bigO_{\PP}\left(\kappa_{n}(\alpha)|\GG_{n}|^{-1/2} \|\widehat{\Pp} - \Pp\|_{L^{2}(\bar{\mu})}^{1 - \gamma/2} \Rr(\widehat{\Pp})^{\gamma/2}\right); \label{eq:case2-1} \\
&\lambda_{n} \,\Rr(\widehat{\Pp})^{2} \leq \bigO_{\PP}\left(\kappa_{n}(\alpha)|\GG_{n}|^{-1/2} \|\widehat{\Pp} - \Pp\|_{L^{2}(\bar{\mu})}^{1 - \gamma/2} \Rr(\widehat{\Pp})^{\gamma/2}\right). \label{case2-2}
\end{align}
Inequality \eqref{eq:case2-1} implies that
\begin{equation}\label{eq:case2-1-1}
\|\widehat{\Pp} - \Pp\|_{L^{2}(\bar{\mu})} \leq \bigO_{\PP}\left( \kappa_{n}(\alpha)^{2/(2+\gamma)} |\GG_{n}|^{-1/(2+\gamma)} \Rr(\widehat{\Pp})^{\gamma/(2+\gamma)} \right).
\end{equation}
Putting \eqref{eq:case2-1-1} into \eqref{case2-2} and using that $\lambda_{n}^{-1} = \smallO_{\PP}( \kappa_{n}(\alpha)^{-4/(2+\gamma)} |\GG_{n}|^{2\mathfrak{e}(\alpha)/(2+\gamma)})$, we get
$
\Rr(\widehat{\Pp})  \leq \smallO_{\PP}(1).
$
Finally, the latter inequality and \eqref{eq:case2-1-1} imply that
\begin{equation*}
\|\widehat{\Pp} - \Pp\|_{L^{2}(\bar{\mu})} \leq \bigO_{\PP}( \kappa_{n}(\alpha)^{2/(2+\gamma)} |\GG_{n}|^{-1/(2 + \gamma)}) \leq \bigO_{\PP}(\lambda_{n}^{1/2}),
\end{equation*}
where the last inequality follows from the fact that $0 < \mathfrak{e}(\alpha) < 1$. 

\bigskip

\noindent {\bf Case \eqref{eq:case3}:}

Inequality \eqref{eq:case3} implies that
\begin{align}
&\|\widehat{\Pp} - \Pp\|_{L^{2}(\bar{\mu})}^{2} \leq \bigO_{\PP} \left(|\GG_{n}|^{-2 \mathfrak{e}(\alpha)/(2+\gamma)}\Rr(\widehat{\Pp}) \right); \label{eq:case3-1} \\
&\lambda_{n} \Rr(\widehat{\Pp})^{2} \leq  \left(|\GG_{n}|^{-2 \mathfrak{e}(\alpha)/(2+\gamma)}\Rr(\widehat{\Pp}) \right). \label{eq:case3-2}
\end{align}

Using that $\lambda_{n}^{-1} = \smallO_{\PP}( \kappa_{n}(\alpha)^{-4/(2+\gamma)} |\GG_{n}|^{2\mathfrak{e}(\alpha)/(2+\gamma)})$, \eqref{eq:case3-2} leads us to
\begin{equation}\label{eq:case3-2-1}
\Rr(\widehat{\Pp}) \leq \bigO_{\PP}(\kappa_{n}(\alpha)^{-4/(2+\gamma)}).
\end{equation}
Putting \eqref{eq:case3-2-1} into \eqref{eq:case3-1}, we get
\begin{equation*}
\|\widehat{\Pp} - \Pp\|_{L^{2}(\bar{\mu})} \leq \bigO_{\PP}(|\GG_{n}|^{-1/(2+\gamma)}) \leq \ \bigO_{\PP}(\lambda_{n}^{1/2}).
\end{equation*}

\noindent {\bf Case \eqref{eq:case1}:}

Finally, from \eqref{eq:case1}, we get $\|\widehat{\Pp} - \Pp\|_{L^{2}(\bar{\mu})} \leq \ \bigO_{\PP}(\lambda_{n}^{1/2})$ and this ends the proof.

\subsection{Proof of Theorem \ref{thm:max-ineq}}
We set
\begin{equation*} 
S = \min\left\{s \geq 0: 2^{-s} \leq \frac{a_{n}}{2^{6} R_{\sigma} \sqrt{|\GG_{n}|}}\right\}.
\end{equation*}
For all $s \in \{0, \ldots, S\}$, we set $H_{s} = \Hh_{B}(2^{-s}R_{\sigma}, \Gg, d_{\Qq}).$ 
Using the definition of $S$, we have
\begin{equation*}
\int_{\frac{a_{n}}{2^{6} \sqrt{|\GG_{n}|}}}^{R_{\sigma}} \Hh_{B}^{1/2}(x, \Gg, d_{Q}) dx \geq \int_{2^{-(S+1)} R_{\sigma}}^{R_{\sigma}} \Hh_{B}^{1/2}(x,\Gg,d_{Q}) dx \geq \frac{1}{2} \sum_{s=0}^{S} 2^{-s} R_{\sigma} H_{s}^{1/2}.
\end{equation*}
Thus the latter inequality and \eqref{eq:cond-an3} implies that
\begin{equation*}
\sum_{s=0}^{S} 2^{-s} R_{\sigma} H_{s}^{1/2} \leq \frac{2 a_{n}}{C_{0} \kappa_{n}(\alpha)} \quad \text{and} \quad \sum_{s=1}^{S} 2^{-s} R_{\sigma} (\sum_{k=0}^{s} H_{k})^{1/2} \leq \frac{4 a_{n}}{C_{0} \kappa_{n}(\alpha)}.
\end{equation*}
For $s \in \{1, \ldots, S\}$, we set
\begin{equation}\label{eq:etaS}
\eta_{s} = \max\left\{ \frac{2^{-3} 2^{-s} (\sum_{k=0}^{s} H_{k})^{1/2} \, C_{0} \, \kappa_{n}(\alpha) \, R_{\sigma}}{a_{n}}; 2^{-(s+3)} \sqrt{s} \right\}.
\end{equation}
Then we have $\sum_{s=1}^{S} \eta_{s} \leq 1.$ 
For $s \in \{0, \ldots, S\}$, we set $N_{s} = N_{B}(2^{-s}R_{\sigma}, \Gg, d_{\Qq})$ and let $\{[\hat{g}_{j_{s}}^{L}, \hat{g}_{j_{s}}^{U}], \\  j_{s} \in \{1, \ldots, N_{s}\}\}$ be a $2^{-s}R_{\sigma}$ recovering with bracketing of $\Gg.$ For $g \in \Gg$, we set $j_{s}^{*} = j_{s}(g) \in \{1, \ldots, N_{s}\}$ such that $g \in [\hat{g}_{j_{s}^{*}}^{L}; \hat{g}_{j_{s}^{*}}^{U}]$ and we define:
\begin{align*}
&g_{j_{s}^{*}}^{U} = \min_{k \leq s} \hat{g}_{j_{k}^{*}}^{U}; \quad g_{j_{s}^{*}}^{L} = \max_{k \leq s} \hat{g}_{j_{k}^{*}}^{L} \quad \text{and} \quad \Delta_{s} = g_{j_{s}^{*}}^{U} - g_{j_{s}^{*}}^{L}  \quad \text{for $s \in \{0, \ldots, S\}$}; \\
&\tau = \min\{s \geq 0: \Delta_{s} \geq K_{s}\} \wedge S \quad \text{with $K_{s} = \frac{2^{4} 2^{-2s} \sqrt{|\GG_{n}|} R_{\sigma}^{2}}{\eta_{s+1} a_{n}}$ for $s \in \{0, \ldots, S - 1\}.$}
\end{align*}
Then we have the following decomposition:
\begin{equation}\label{eq:decom-g}
g = g_{j_{0}^{*}}^{L} + \sum_{s=0}^{S} (g - g_{j_{s}^{*}}^{L}) \ind_{\{\tau = s\}} + \sum_{s = 1}^{S} (g_{j_{s}^{*}}^{L} - g_{j_{s-1}^{*}}^{L}) \ind_{\{\tau \geq s\}}.
\end{equation}
For $g \in \Gg$, we set $\widetilde{g} = g - \langle \mu,g \rangle.$ Using \eqref{eq:decom-g}, we have
\begin{equation*}
\PP\left( \left| \frac{1}{\sqrt{|\GG_{n}|}} \sum_{u \in \GG_{n}} \widetilde{g}(X_{u}) \right| > a_{n} \, \, \text{for some $g \in \Gg$} \right) \leq \pmb{\PP_{I}} + \pmb{\PP_{II}} + \pmb{\PP_{III}}, 
\end{equation*}
where
\begin{align*}
&\pmb{\PP_{I}} = \PP\left( \frac{1}{\sqrt{|\GG_{n}|}} \left| \sum_{u \in \GG_{n}} \widetilde{g}_{j_{0}^{*}}^{L}(X_{u}) \right| > \frac{a_{n}}{4} \, \, \text{for some $g \in \Gg$} \right) \\
&\pmb{\PP_{II}} = \PP\left( \frac{1}{\sqrt{|\GG_{n}|}} \left| \sum_{s=0}^{S} \sum_{u \in \GG_{n}}  (\widetilde{g} - \widetilde{g}_{j_{s}^{*}}^{L})(X_{u}) \ind_{\{\tau = s\}} \right| > \frac{a_{n}}{2} \, \, \text{for some $g \in \Gg$} \right) \\  
&\pmb{\PP_{III}} = \PP\left( \frac{1}{\sqrt{|\GG_{n}|}} \left| \sum_{s=1}^{S} \sum_{u \in \GG_{n}}  (\widetilde{g}_{j_{s}^{*}}^{L} - \widetilde{g}_{j_{s - 1}^{*}}^{L})(X_{u}) \ind_{\{\tau \geq s\}} \right| > \frac{a_{n}}{4} \, \, \text{for some $g \in \Gg$} \right). 
\end{align*}
Now, we will bound each term on the right hand side of the previous inequality. In the sequel, we will denote by $c$ a generic constant which may vary line to line. We replace $C$ given in Lemma \ref{lem:bernstein} by $c_{1}$ and $c_{2}$ will denote a constant which depend on $c$, $C_{0}$, $C$ and $C_{1}$.
\subsection*{Bound of $ \pmb{\PP_{I}}$}

Using Lemma \ref{lem:bernstein}, we have
\begin{equation}\label{eq:ineq-H0}
\PP\left( \frac{1}{\sqrt{|\GG_{n}|}} \left| \sum_{u \in \GG_{n}} \widetilde{g}_{j_{0}^{*}}^{L}(X_{u}) \right| > \frac{a_{n}}{4} \, \, \text{for some $g \in \Gg$} \right) \leq c_{1} \exp\left( H_{0} - \frac{4 a_{n}^{2}}{c(K  a_{n}  v_{n}(\alpha) + 4  R_{\sigma}^{2})}\right).
\end{equation}
Next, using \eqref{eq:cond-an3}, we have 
\begin{equation}\label{eq:H0-an}
a_{n}  > \frac{C_{0} \, \kappa_{n}(\alpha) \, R_{\sigma} H_{0}^{1/2}}{2} \quad \text{and then that} \quad H_{0} \leq \frac{4 \, a_{n}^{2}}{C_{0}^{2} \, \kappa_{n}(\alpha)^{2} \, R_{\sigma}^{2}}.
\end{equation}
Using \eqref{eq:H0-an} in \eqref{eq:ineq-H0} and $v_{n}(\alpha) < \kappa_{n}(\alpha)$, it then follows that
\begin{equation*}
\PP\left( \frac{1}{\sqrt{|\GG_{n}|}} \left| \sum_{u \in \GG_{n}} \widetilde{g}_{j_{0}^{*}}^{L}(X_{u}) \right| > \frac{a_{n}}{4} \, \, \text{for some $g \in \Gg$} \right) \leq c_{1}  \exp\left(\frac{4 \, a_{n}^{2}}{C_{0}^{2} \, \kappa_{n}(\alpha)^{2} \, R_{\sigma}^{2}} - \frac{4 a_{n}^{2}}{c \, \kappa_{n}(\alpha)^{2} \, (K a_{n} + 4 R_{\sigma}^{2})}\right).
\end{equation*}
Using \eqref{eq:cond-an1} and \eqref{eq:cond-C0-C} and taking $C$ large enough in the last inequality, we get
\begin{align*}
\PP\left( \frac{1}{\sqrt{|\GG_{n}|}} \left| \sum_{u \in \GG_{n}} \widetilde{g}_{j_{0}^{*}}^{L}(X_{u}) \right| > \frac{a_{n}}{4} \, \, \text{for some $g \in \Gg$} \right) &\leq c_{1} \exp\left( \frac{4 a_{n}^{2} }{C^{2} (C_{1} + 4) \kappa_{n}(\alpha)^{2} \,R_{\sigma}^{2}} - \frac{4 a_{n}^{2}}{c  (C_{1} + 4) \kappa_{n}(\alpha)^{2} R_{\sigma}^{2}}\right) \\ 
&\leq c_{1} \exp\left( - \frac{c_{2} a_{n}^{2}}{R_{\sigma}^{2} \, \kappa_{n}(\alpha)^{2}} \right).
\end{align*}

\subsection*{Bound of $\pmb{\PP_{II}}$}

From the definition of $\tau$, we have
\begin{align}
\label{eq:A7} &\Delta_{s} \ind_{\{\tau = s\}} \geq K_{s} \quad \forall s \in \{0, \ldots, S-1\} \quad \text{and} \\
\label{eq:A7bis} &\Delta_{s} \ind_{\{\tau = s\}} \leq \Delta_{s-1} \ind_{\{\tau = s\}} \leq K_{s-1} \quad \forall s \in \{1, \ldots, S\}. 
\end{align}
It follows from the definition of bracketing that
\begin{equation}\label{eq:A10}
0 \leq g - g_{j_{s}^{*}} \leq \Delta_{s} \quad \forall s \in \{0, \ldots, S\}.
\end{equation}
Using \eqref{eq:A7} in the first inequality, the fact $\mu\Qq= \mu$ in the first equality and the definition of the bracketing in the third inequality, we get, for $s \in \{0, \ldots, S-1\}$,
\begin{multline}\label{eq:A10-s}
\frac{1}{|\GG_{n}|} \sum_{u \in \GG_{n}} \langle \mu, \Delta_{s} \ind_{\{\tau = s\}} \rangle \leq \frac{1}{K_{s}} \langle \mu, \Delta_{s}^{2} \rangle = \frac{1}{K_{s}} \langle \mu, \Qq(\Delta_{s}^{2}) \rangle \\ 
\leq \frac{d^{2}_{\Qq}(g_{j_{s}^{*}}^{L},g_{j_{s}^{*}}^{U})}{K_{s}} \leq \frac{2^{-2s} R_{\sigma}^{2}}{K_{s}} = \frac{\eta_{s+1} a_{n}}{2^{4} \sqrt{|\GG_{n}|}}.
\end{multline}
Using the definition of $S$ and the previous arguments, we have
\begin{multline}\label{eq:A10-t}
\frac{1}{|\GG_{n}|} \sum_{u \in \GG_{n}} \langle \mu, \Delta_{S}\ind_{\{\tau = S\}} \rangle \leq (\langle \mu,\Delta_{S}^{2} \rangle)^{1/2} = (\langle \mu,\Qq(\Delta_{S}^{2}) \rangle)^{1/2} \leq d_{\Qq}(g_{j_{s}^{*}}^{L},g_{j_{s}^{*}}^{U}) \\ 
\leq 2^{-S} R_{\sigma} \leq \frac{a_{n}}{2^{4} \sqrt{|\GG_{n}|}}.
\end{multline}
Using \eqref{eq:A10-s}, \eqref{eq:A10-t} and the fact that $\sum_{s=0}^{S} \eta_{s} \leq 1$, we get
\begin{equation}\label{eq:A10-q}
\sum_{s=0}^{S} \frac{1}{\sqrt{|\GG_{n}|}} \sum_{u \in \GG_{n}} \langle \mu, \Delta_{s}\rangle \ind_{\{\tau = s\}} \leq \frac{a_{n}}{2^{4}} \sum_{s = 0}^{S-1} \eta_{s+1} + \frac{a_{n}}{2^{4}} \leq \frac{a_{n}}{2^{3}}.
\end{equation}
It follows from \eqref{eq:A10} and \eqref{eq:A10-q} that
\begin{align*}
\left|\sum_{s=0}^{S} \frac{1}{\sqrt{|\GG_{n}|}} \sum_{u \in \GG_{n}} (\widetilde{g} - \widetilde{g}_{j_{s}^{*}}^{L})(X_{u}) \ind_{\{\tau = s\}}\right| & \leq \left| \sum_{s = 0}^{S} \frac{1}{\sqrt{|\GG_{n}|}} \sum_{u \in \GG_{n}} (\Delta_{s} - \langle \mu, g - g_{j_{s}^{*}}^{L} \rangle)(X_{u}) \ind_{\{\tau = s\}}\right| \\
& \leq \left| \sum_{s = 0}^{S} \frac{1}{\sqrt{|\GG_{n}|}} \sum_{u \in \GG_{n}} (\Delta_{s} - \langle \mu, \Delta_{s} \rangle)(X_{u}) \ind_{\{\tau = s\}}\right| \\ 
& \hspace{3cm} +  \sum_{s = 0}^{S} \frac{1}{\sqrt{|\GG_{n}|}} \sum_{u \in \GG_{n}} \langle \mu, \Delta_{s} \rangle \ind_{\{\tau = s\}} \\
& \leq \left| \sum_{s = 0}^{S} \frac{1}{\sqrt{|\GG_{n}|}} \sum_{u \in \GG_{n}} (\Delta_{s} - \langle \mu, \Delta_{s} \rangle)(X_{u}) \ind_{\{\tau = s\}}\right| + \frac{a_{n}}{2^{3}}.
\end{align*}
This implies
\begin{equation*}
\pmb{\PP_{II}} \leq \PP\left( \left| \sum_{s = 0}^{S} \frac{1}{\sqrt{|\GG_{n}|}} \sum_{u \in \GG_{n}} (\Delta_{s} - \langle \mu, \Delta_{s} \rangle)(X_{u}) \ind_{\{\tau = s\}}\right| \geq \frac{a_{n}}{2} - \frac{a_{n}}{2^{3}} \quad \text{for some $g \in \Gg$} \right) \leq \pmb{\PP_{II}^{(1)}} + \pmb{\PP_{II}^{(2)}},
\end{equation*}

where
\begin{align*}
&\pmb{\PP_{II}^{(1)}} = \PP\left(  \left|\frac{1}{\sqrt{|\GG_{n}|}} \sum_{u \in \GG_{n}}  (\widetilde{g}_{j_{0}^{*}}^{U} - \widetilde{g}_{j_{0}^{*}}^{L})(X_{u}) \ind_{\{\tau = 0\}} \right| > \frac{a_{n}}{2^{3}} \, \, \text{for some $g \in \Gg$} \right); \\
&\pmb{\PP_{II}^{(2)}} = \PP\left( \left| \sum_{s = 1}^{S} \frac{1}{\sqrt {|\GG_{n}|}} \sum_{u \in \GG_{n}} (\widetilde{g}_{j_{s}^{*}}^{U} - \widetilde{g}_{j_{s}^{*}}^{L})(X_{u}) \ind_{\{\tau = s\}}\right| > \frac{a_{n}}{4} \quad \text{for some $g \in \Gg$}  \right).
\end{align*}
On the one hand,  note that we have
\begin{equation*}
\|\Qq((g_{j_{0}^{*}}^{U} - g_{j_{0}^{*}}^{L})^{2})\|_{\infty} \leq R_{\sigma}^{2} \quad \text{and} \quad c_{2}(g_{j_{0}^{*}}^{U} - g_{j_{0}^{*}}^{L}) \leq 2 \, K
\end{equation*}
in such a way that using Lemma \ref{lem:bernstein}, \eqref{eq:H0-an}, the fact that $v_{n}(\alpha) \leq \kappa_{n}(\alpha)^{2}$ and taking $C$ large enough in \eqref{eq:cond-C0-C}, we get
\begin{align*}
\pmb{\PP_{II}^{(1)}} &\leq c_{1} \exp\left( H_{0} - \frac{4 a_{n}^{2}}{c(K v_{n}(\alpha) a_{n} + 4 R_{\sigma^{2}})} \right) \\ 
& \leq c_{1} \exp\left( \frac{4 a_{n}^{2}}{C_{0}^{2} \kappa_{n}(\alpha)^{2} R_{\sigma}^{2}} - \frac{4 a_{n}^{2}}{c \kappa_{n}(\alpha)^{2} R_{\sigma}^{2}(C_{1} + 4)} \right) \\ 
&\leq c_{1} \exp\left( - \frac{c_{2} a_{n}^{2}}{R_{\sigma}^{2} \kappa_{n}(\alpha)^{2}}\right).
\end{align*}
On the other hand, using the definition of bracketing, \eqref{eq:A7bis} and \eqref{eq:etaS}, we have
\begin{equation}\label{eq:A7ter}
\|\Qq(\Delta_{s}^{2})\|_{\infty} \leq 2^{-2s} R_{\sigma}^{2}; \quad c_{2}(\Delta_{s}) \leq 2 K_{s-1} \quad \text{and} \quad \sum_{k=0}^{s} H_{k} \leq \frac{2^{6} a_{n}^{2} \eta_{s}^{2} }{2^{-2s} \, C_{0}^{2} \, R_{\sigma}^{2} \, \kappa_{n}(\alpha)^{2}}.   
\end{equation}
From the definition of $K_{s}$ and $\eta_{s}$, one can check that
\begin{equation}\label{eq:A7quart}
K_{s-1} v_{n}(\alpha) a_{n} \eta_{s} = 2^{6} 2^{-2s} R_{\sigma}^{2} \kappa_{n}(\alpha)^{2}.
\end{equation}
Then, using \eqref{eq:A7ter}, Lemma \eqref{lem:bernstein}, \eqref{eq:A7quart}, \eqref{eq:cond-C0-C} for $C$ large enough and \eqref{eq:etaS}, we get
\begin{align*}
\pmb{\PP_{II}^{(2)}} &\leq \sum_{s=1}^{S} \PP\left( \left| \frac{1}{\sqrt{|\GG_{n}|}} \sum_{u \in \GG_{n}} (\Delta_{s} - \langle \mu,\Delta_{s} \rangle)(X_{u}) \ind_{\{\tau = s\}} \right| > \frac{a_{n} \eta_{s}}{4} \quad \text{for some $g \in \Gg$} \right) \\
&\leq \sum_{s=1}^{S} c_{1} \exp\left( \sum_{k \leq s} H_{k} - \frac{2 a_{n}^{2} \eta_{s}^{2}}{c(K_{s-1} a_{n} v_{n}(\alpha) \eta_{s} +  2\times2^{-2s} R_{\sigma}^{2})} \right) \\
& \leq \sum_{s=1}^{S} c_{1} \exp\left( \frac{2^{6} a_{n}^{2} \eta_{s}^{2}}{2^{-2s} \, R_{\sigma}^{2} \, C_{0}^{2} \, \kappa_{n}(\alpha)^{2}} - \frac{a_{n}^{2} \eta_{s}^{2}}{c \, 2^{-2s} R_{\sigma}^{2} \, \kappa_{n}(\alpha)^{2}}\right) \\
& \leq c_{1} \sum_{s=1}^{S} \exp\left( - \frac{a_{n}^{2}  \eta_{s}^{2}}{c 2^{-2s} R_{\sigma}^{2}} \right) \leq c_{1} \sum_{s=1}^{S} \exp\left( - \frac{a_{n}^{2}  s}{c \, R_{\sigma}^{2} \, \kappa_{n}(\alpha)^{2}} \right) \leq c_{1} \exp\left( - \frac{c_{2} a_{n}^{2} }{R_{\sigma}^{2} \kappa_{n}(\alpha)^{2}} \right).
\end{align*}

\subsection*{Bound of $\pmb{\PP_{III}}$}

We have
\begin{equation*}
0 \leq g_{j_{s}^{*}}^{L} - g_{j_{s-1}^{*}}^{L} \leq g_{j_{s-1}^{*}}^{U} - g_{j_{s-1}^{*}}^{L} = \Delta_{s-1} \quad \text{and} \quad \Delta_{s-1} \ind_{\{\tau \geq s\}} \leq K_{s-1}.
\end{equation*}
This implies that
\begin{align*}
&\|\Qq(g_{j_{s}^{*}}^{L} - g_{j_{s-1}^{*}}^{L})^{2} \ind_{\{\tau \geq s\}}\|_{\infty} \leq \|\Qq(g_{j_{s-1}^{*}}^{U} - g_{j_{s-1}^{*}}^{L})^{2}\|_{\infty} \leq 2^{-2(s-1)} R_{\sigma}^{2} \quad \text{and} \\ 
&c_{2}((g_{j_{s}^{*}}^{L} - g_{j_{s-1}^{*}}^{L})\ind_{\{\tau \geq s\}}) \leq \frac{4}{3} (1 + R \alpha) \|\Delta_{s-1} \ind_{\{\tau \geq s\}}\|_{\infty} \leq \frac{4}{3} (1 + R \alpha) \frac{2^{6} 2^{-2s}}{\eta_{s} a_{n}}.
\end{align*}
Now, following similar arguments as in the bound of $\pmb{\PP_{II}^{(2)}}$, we are led to
\begin{equation*}
\pmb{\PP_{III}} \leq  c_{1} \exp\left(- \frac{c_{2} \, a_{n}^{2}}{R_{\sigma}^{2} \kappa_{n}(\alpha)^{2}} \right).
\end{equation*}  

\subsection{Proof of Lemma \ref{lem:modulus-c}}$\,$

\medskip

First we treat \eqref{eq:modulus-c1}. We choose $C_{0}$ large enough and we fix $A_{0}$ and $K$. We choose a positive constant $c_{r}$ and we set
\begin{equation*}
C_{1} = \frac{K C_{0} A_{0}}{c_{r}} \quad \text{and} \quad c_{a} = \frac{c_{r}^{2} C_{1}}{K}.
\end{equation*}
Then, setting
\begin{equation*}
R_{\sigma} = c_{r} \delta_{n} \quad \text{and} \quad a_{n} = c_{a} |\GG_{n}^{1/2}| \delta_{n}^{2},
\end{equation*}
one can check that conditions \eqref{eq:cond-an1}-\eqref{eq:cond-C0-C} are satisfied. We can therefore apply Theorem \ref{thm:max-ineq} and this leads us to inequality \eqref{eq:modulus-c1}.

\medskip

Now we treat  \eqref{eq:modulus-c2}. We set
\begin{equation*}
S = \min\left\{ k > 1: 2^{-k} < \delta_{n} \right\}.
\end{equation*}
In the sequel, we will denote by $c$ a generic constant  independent of $s \in \{1, \ldots, S\}$ and which may varies line to line. We have
\begin{multline}\label{eq:peeling}
\PP\left( \sup_{g \in \Gg; \, d_{\Qq}(g,g_{0}) \, >  \, \delta_{n}} \frac{\left| \frac{1}{\sqrt{|\GG_{n}|}} \sum_{u \in \GG_{n}} (\widetilde{g} - \widetilde{g}_{0})(X_{u}) \right|}{d_{\Qq}(g,g_{0})^{1 - \gamma/2} } \geq \kappa_{n}(\alpha) T\right) \\ 
\leq \sum_{s=1}^{S} \PP\left( \sup_{g \in \Gg(2^{-s+1})} \left| \frac{1}{\sqrt{|\GG_{n}|}} M_{\GG_{n}}(\widetilde{g} - \widetilde{g}_{0}) \right| \geq T \kappa_{n}(\alpha) \left(2^{-s+1}\right)^{1-\gamma/2} \right).
\end{multline}
Now let $s \in \{1, \ldots, S\}$. We set $a_{n} = T \kappa_{n}(\alpha) (2^{-s+1})^{1-\gamma/2}$ and $R_{\sigma} = c \, 2^{-s+1}$. Then, with a good choice of the constants, one can check the conditions \eqref{eq:cond-an1}-\eqref{eq:cond-C0-C} in such a way that we have
\begin{equation*}
\PP\left( \sup_{g \in \Gg(2^{-s+1})} \left| \frac{1}{\sqrt{|\GG_{n}|}} M_{\GG_{n}}(\widetilde{g} - \widetilde{g}_{0}) \right| \geq T \kappa_{n}(\alpha) \left(2^{-s+1}\right)^{1-\gamma/2} \right) \leq c \, \exp\left( - c T  \left( 2^{-s+1} \right)^{-\gamma} \right).
\end{equation*}
Putting the latter inequality into \eqref{eq:peeling}, we get inequality  \eqref{eq:modulus-c2}.

\bibliographystyle{abbrv}
\bibliography{biblio}
\end{document}